\documentclass[aps, prd, amsmath, floats, floatfix, nofootinbib, superscriptaddress,twocolumn]{revtex4-1}
\usepackage{graphicx}
\usepackage{color}
\usepackage{latexsym}
\usepackage{epsfig}
\usepackage[colorlinks]{hyperref}
\usepackage{array}
\usepackage{enumitem}
\usepackage{textcomp}
\usepackage{footmisc}
\usepackage{xfrac}
\usepackage[utf8]{inputenc}

\newcommand{\be}{\begin{eqnarray}}
\newcommand{\ee}{\end{eqnarray}}
\newcommand{\bead}{\begin{eqnarray}\begin{aligned}}
\newcommand{\eead}{\end{aligned}\end{eqnarray}}
\newcommand{\beq}{\begin{equation*}}
\newcommand{\eeq}{\end{equation*}}

\newcommand{\bitem}{\begin{itemize}}
\newcommand{\eitem}{\end{itemize}}
\newcommand{\bwide}{\begin{widetext}}
\newcommand{\ewide}{\end{widetext}}

\usepackage{subcaption}

\begin{document}

\title{GRMHD beyond Kerr: An extension of the HARM code for thin disks to non-Kerr spacetimes}
\author{Sourabh Nampalliwar} 
\email[Corresponding author: ]{sourabh.nampalliwar@uni-tuebingen.de}
\affiliation{Theoretical Astrophysics, IAAT, Eberhard Karls Universit\"{a}t, T\"{u}bingen, Germany}
\affiliation{SISSA, Via Bonomea 265, 34136 Trieste, Italy and INFN Sezione di Trieste}
\affiliation{IFPU - Institute for Fundamental Physics of the Universe, Via Beirut 2, 34014 Trieste, Italy}
\author{Aristomenis I. Yfantis}
\affiliation{Theoretical Astrophysics, IAAT, Eberhard Karls Universit\"{a}t, T\"{u}bingen, Germany}
\affiliation{Department of Astrophysics/IMAPP, Radboud University, P.O. Box 9010, 6500 GL Nijmegen, The Netherlands}
\author{Kostas D. Kokkotas}
\affiliation{Theoretical Astrophysics, IAAT, Eberhard Karls Universit\"{a}t, T\"{u}bingen, Germany}

\begin{abstract}
    Black hole based tests of general relativity have proliferated in recent times with new and improved detectors and telescopes. Modelling of the black hole neighborhood, where most of the radiation carrying strong-field signature originates, is of utmost importance for robust and accurate constraints on possible violations of general relativity. As a first step, this paper presents the extension of general relativistic magnetohydrodynamic simulations of thin accretion disks to parametrically deformed black holes that generalize the Kerr solution. The extension is based on \textsc{harmpi}, a publicly available member of the \textsc{harm} family of codes, and uses a phenomenological metric to study parametric deviations away from Kerr. The extended model is used to study the disk structure, stability, and radiative efficiency. We also compute the Fe K$\alpha$ profiles in simplified scenarios and present an outlook for the future.
\end{abstract}
\date{\today}

\maketitle

\section{Introduction}
%Introduction to testing GR
Tests of theories of gravity in the strong-field regime have become an important field of research in the past few years with the detection of gravitational waves and advances in x-ray spectroscopy, relativistic astrometry, and black hole imaging. While general relativity (GR) has been hugely successful in the weak-field regime~\cite{Will2014}, its validity in the strong-field regime has recently become testable ~\cite{LIGOScientific:2021sio,Tripathi:2020yts,GRAVITY:2018ofz}, and these tests are expected to get more sensitive in the coming decades. 

%Why use BHs to test GR?
One of the most remarkable predictions of GR is the existence of black holes (BHs). BHs, being extremely compact, give rise to the strongest gravitational fields in our universe. They can be completely described, at least within GR, with just a few parameters (due to the no-hair theorem~\cite{Israel:1967wq,Israel:1967za,Hawking:1971vc,Robinson:1975bv}). Moreover, many of them can be found in our universe, with estimates placing roughly a million of them within our own galaxy~\cite{Elbert:2017sbr}. These properties make BHs the ideal candidates for performing strong-field tests of GR.

%\co{Add a paragraph about x-ray spectroscopy} 
BH-based tests of gravity in the strong-field regime have become accessible with a variety of observational techniques in recent years. Gravitational waves-based tests began in 2016~\cite{LIGOScientific:2016lio} (gravitational wave-based tests using pulsar timing have been around for a few decades~\cite{Taylor:1982zz}, but these do not probe the strong-field regime as closely as the other listed techniques, see Fig.~2 in Ref.~\cite{Psaltis:2008bb}), astrometry-based tests of GR from the center of the Milky Way were reported in 2018~\cite{GRAVITY:2018ofz}; in the same year x-rays were first used to probe deviations from GR~\cite{Cao2017}, and BH shadow-based tests were first performed in 2019~\cite{EventHorizonTelescope:2019ggy}. X-ray spectroscopy, in particular, is one of the most interesting and promising techniques in this regard. It is based on the idea of using radiation emanating from and/or traveling close to the BH and thus carrying imprint of its nature. Feasible with both stellar-mass and supermassive BHs alike, with several astrophysical sources already observed, and with relatively cleaner signal than some other techniques, x-ray reflection spectroscopy provides a versatile, independent and complimentary tool to study theories of gravity.

%Why the BH neighborhood is important?
When performing strong-field tests, the immediate neighborhood of the BH is an especially important region, since many of the tests are based on analyzing radiation emanating from here. Precise and accurate tests are, therefore, predicated on precise and accurate modelling of the BH neighborhood. Such modelling is, however, fraught with difficulties. The first serious difficulty is the handling of the alternative theory and the black hole metric being tested, the latter of which can have a complex numerical, or even unknown, form. The second serious difficulty is the handling of complex partial differential equations describing the intermixing of gravitational, electromagnetic and kinetic effects in the BH neighborhood, manifesting even for GR and around Kerr BHs. 

In light of the above difficulties in modelling the BH neighborhood, recent years have seen attempts to constrain alternative theories of gravity and possible violations of GR using the approach of decreased fidelity: most of these studies are based on idealistic models of the BH neighborhood, assuming, for instance, an infinitesimally thin accretion disk, perfectly circular orbits for the accreting matter, and so on. Apart from introducing extra parameters in the model, which gives rise to irreducible degeneracies, these studies rely on the fundamental assumption that theoretical models of the BH neighborhood, developed assuming the validity of (and/or compatibility with) GR, continue to remain valid in non-GR scenarios that these studies set out to investigate. This makes the whole analysis, at best, unreliable and, at worst, misleading. It is, therefore, imperative to evolve the models of BH neighborhoods in non-GR scenarios to a robust, accurate and reliable framework that can be used to perform accurate tests of GR.

The most promising approach to achieve high-fidelity modelling of the BH neighborhood is magnetohydrodynamic (MHD) simulations. Significant progress has been made in this regard within GR. GRMHD simulations solve coupled systems of partial differential equations comprising of magneto-hydrodynamic equations and gravitational field equations. In generic settings (e.g., when simulating gravitational collapse into a BH), one needs to evolve both the \textit{fluid} and the \textit{background} gravitational field. However, accretion of matter can often be modeled in a relatively simpler manner by treating the background field as fixed.\footnote{This can also be understood by realizing that the accretion rate is so small that the mass, and other parameters of the BH, do not change significantly over the timescale of the evolution.} These \textit{test-fluid} GRMHD simulations have successfully been applied to diverse astrophysical scenarios, especially when combined with high-resolution shock-capturing (HRSC) techniques. (See Ref.~\cite{Font:2008fka} for other techniques like artificial viscosity, smooth particle hydrodynamics, etc.) One of the most popular test-fluid HRSC GRMHD codes in the market today is \textsc{harm}~\cite{Gammie:2003rj,Noble:2005gf}. It uses an energy-conservative method to solve equations of ideal MHD in the Kerr-Schild coordinates. Various extensions of \textsc{harm} have been developed, to handle three spatial dimensions~\cite{Noble:2008tm}, parallelization~\cite{Tchekhovskoy2019}, M1 radiation closure~\cite{McKinney:2013txa}, GPU-acceleration~\cite{Liska:2019uqw}, and so on. Another popular choice is the \textit{black hole accretion code} (BHAC)~\cite{Porth:2016rfi}. It uses adaptive mesh refinement techniques to efficiently model the neighborhood, and has been used to simulate images of non-GR BHs~\cite{Olivares:2018abq,Mizuno:2018lxz}.

%\co{Details.} 
%\co{Say a few words about top-down and bottom-up metrics, and why we chose to go with the latter in this work.}
Modelling of the BH itself can be approached in two ways. In the first approach, one begins with a specific alternative theory of gravity and uses the BH solutions of that particular theory. In the second approach, one begins with a generic metric and parametrizes it relative to the Kerr metric (often by imposing one or more symmetries that the Kerr metric possesses). Each approach has its advantages and disadvantages. While using a theory-specific metric provides a direct handle on the validity of a physical theory, finding BH solutions in alternative theories of gravity, as in GR, can be extremely difficult. Writing the solution in a form that is convenient for computation adds to the complication. Theory-agnostic metrics can often be written down in a relatively convenient form, but do not have a physical theory behind them and can be difficult to interpret. They can, however, be used without reservation for \textit{null} tests of GR, where parametrized deviations from the BH solutions of GR (often the rotating one known as the Kerr solution~\cite{Kerr1963}) provide a handle on violations of GR via the no-hair theorem. 

Within the simplifying assumptions (infinitesimally thin disk with matter in circular orbits, etc.~\cite{Bambi:2016sac}), some of us have developed a framework, \textsc{relxill\_nk}, to test alternative BHs (of both theory-specific and theory-agnostic types) using x-rays~\cite{Bambi:2016sac,Abdikamalov:2019yrr}. \textsc{relxill\_nk} has been used to test for deviations from the Kerr solution in both stellar-mass~\cite{Zhang:2019zsn,Xu2018,Tripathi:2020yts,Yu:2021xen} and supermassive BHs~\cite{Cao2017,Choudhury:2018zmf,Tripathi2018a,Nampalliwar:2019iti,Abdikamalov:2021zwv}, and it has provided some of the best constraints so far on such deviations. 
In this work, we take this framework forward and describe our first step towards high-fidelity tests of gravity using x-ray reflection spectroscopy. We present our attempt to model the neighborhood of a non-GR BH described by a theory-agnostic metric using the public, parallelized three-dimensional version of \textsc{harm}, namely, \textsc{harmpi}~\cite{Tchekhovskoy2019}. We mostly perform the simulations in two dimensions (radial and polar, assuming symmetry in the azimuthal direction) due to limitations on computational cost, but show one 3D simulation to illustrate that the qualitative results from 2D simulations hold. This simplified case serves as a stepping stone towards our ultimate goal of high-fidelity modelling of the BH neighborhood. We focus on thin disks, i.e., disks with the Eddington accretion rate in the range $0.1-0.3$, since x-ray reflection spectroscopy has primarily been used for such systems and numerical simulations in GR have shown good agreement with models used for data analysis (which use analytic infinitesimally-thin disks).  We evolve the classic tori profiles described in Ref. \cite{1985ApJ...288....1C} and calculate several properties of the final disk related to stability and radiation. In particular, we compute the iron line profile for MHD-modeled systems and compare them with those for analytic thin-disk systems, finding good agreement. 

%\subsection{Executive summary}

%Outline
The rest of the paper is organized as follows. Sec.~\ref{sec:review} presents a review of the theory-agnostic metric used to model the BH, and \textsc{harmpi}. In Sec.~\ref{sec:model}, we describe our enhancements to \textsc{harmpi}. Simulation details are given in Sec.~\ref{sec:simulation}. In Sec.~\ref{sec:results} we present the results ranging from disk structure and stability to radiative efficiency and spectrum. We end with Sec.~\ref{sec:future} with an outlook for the future. In what follows, greek letters are used for spacetime indices, and roman for purely spatial indices. The signature of the metric is $(-+++)$ and geometrized units are used such that $G=c=1$.
  
\section{Review\label{sec:review}}

\subsection{Johannsen metric\label{subsec:metric}} 
One of the most popular choices of theory-agnostic non-Kerr metrics is the one proposed in Ref.~\cite{Johannsen2015} (Johannsen metric hereafter). It preserves the separability of Hamilton-Jacobi equations of the Kerr solution (which leads to the existence of a constant of motion, known as the Carter constant in the Kerr case~\cite{Carter1971}, and can be traced back to the conservation of the Killing-Yano tensor~\cite{Papadopoulos:2018nvd}) while introducing four new radial functions that quantify the deviation of the metric from Kerr. In Boyer-Lindquist coordinates, the metric components are given as~\cite{Johannsen2015}
\be
g_{tt}^{\rm BL} &=& - \frac{\tilde{\Sigma} \left(\Delta - a^2 A_2^2 \sin^2\theta\right)}{B^2}, \nonumber \\
g_{t\phi}^{\rm BL} &=& - \frac{2 a \tilde{\Sigma} \sin^2\theta}{B^2} \left[ \left(r^2 + a^2\right) A_1 A_2 - \Delta\right], \nonumber \\
g_{rr}^{\rm BL} &=& \frac{\tilde{\Sigma}}{\Delta A_5}  \nonumber \\
g_{\theta\theta}^{\rm BL} &=& \tilde{\Sigma}, \nonumber \\
g_{\phi\phi}^{\rm BL} &=& \frac{\tilde{\Sigma} \sin^2\theta}{B^2} \left[ \left(r^2 + a^2\right)^2 A_1^2 - a^2 \Delta \sin^2\theta\right],\nonumber
\label{eq:metric_BL}
\ee
where the superscript BL indicates that the components are given in Boyer-Lindquist coordinates, $M$ is the BH mass, $a = J/M$ is the BH spin,
\be
&&B = \left(r^2 + a^2\right) A_1 - a^2 A_2 \sin^2\theta \, , \quad
\tilde{\Sigma} = \Sigma + f \, , \nonumber\\
&&\Sigma = r^2 + a^2 \cos^2\theta \, , \quad
\Delta = r^2 - 2 M r + a^2 \, ,\nonumber
\ee
and $\{f, A_1, A_2, A_5\}$ are the four deviation functions. The deviation functions are written in terms of power series over $M/r$, and after setting the lower order coefficients, dictating far-field behavior of the spacetime, to their GR values, and retaining the leading order parameters that dictate near-field behavior of the spacetime, we get

\be\label{eq-def}
&& f = \epsilon_3 \frac{M^3}{r} \, , \qquad
A_1 = 1 + \alpha_{13} \left(\frac{M}{r}\right)^3 \, , \nonumber\\
&& A_2 = 1 + \alpha_{22} \left(\frac{M}{r}\right)^2 \, , \qquad
A_5 = 1 + \alpha_{52} \left(\frac{M}{r}\right)^2 \, .\nonumber
\ee
%\end{widetext}
Here, $\epsilon_3$, $\alpha_{13}$, $\alpha_{22}$, and $\alpha_{52}$ are the leading order deviation parameters and are dimensionless. %Such a metric has the correct Newtonian limit and is consistent with the current PPN constraints~\cite{j-m}. It 
The metric reduces to the Kerr metric for $\epsilon_3 = \alpha_{13} = \alpha_{22} = \alpha_{52} = 0$.

Since the \textsc{harm} family of codes also require the metric in horizon-penetrating coordinates, we use the above metric in Kerr-Schild-like coordinates, also given in Ref.~\cite{Johannsen2015}:
\be
g_{tt}^{\rm KS} &=& -\frac{\tilde{\Sigma}[\Delta-a^2A_2(r)^2\sin^2\theta]}{F}, \nonumber \\
g_{tr}^{\rm KS} &=& \frac{\tilde{\Sigma}}{\sqrt{A_5(r)}F} \{A_1(r)[2Mr+a^2A_2(r)^2\sin^2\theta] \nonumber \\
&& - a^2A_2(r)\sin^2\theta\}, \nonumber \\
g_{t\phi}^{\rm KS} &=& -\frac{a\tilde{\Sigma}[(r^2+a^2)A_1(r)A_2(r)-\Delta]\sin^2\theta}{F}, \nonumber \\
g_{rr}^{\rm KS} &=& \frac{\tilde{\Sigma}A_1(r)}{A_5(r)F} \{ A_1(r)[\Delta+4Mr+a^2A_2(r)^2\sin^2\theta] \nonumber \\
&& - 2a^2A_2(r)\sin^2\theta \}, \nonumber \\
g_{r\phi}^{\rm KS} &=& -\frac{a\tilde{\Sigma}\sin^2\theta }{ \sqrt{A_5(r)}F } [(r^2+a^2)A_1(r)^2A_2(r) \nonumber \\
&& +2MrA_1(r)-a^2A_2(r)\sin^2\theta], \nonumber \\
g_{\theta\theta}^{\rm KS} &=& \tilde{\Sigma}, \nonumber \\
g_{\phi\phi}^{\rm KS} &=& \frac{\tilde{\Sigma} \left[(r^2+a^2)^2A_1(r)^2-a^2\Delta\sin^2\theta\right]\sin^2\theta }{ F },
\label{eq:metric_KS}
\ee
where
\beq
F \equiv \left[ (r^2+a^2)A_1(r)-a^2A_2(r)\sin^2\theta \right]^2.
\eeq

%The Johannsen metric is a popular choice for performing parameterized tests of gravity, and has been studied using gravitational waves~\cite{}, shadows~\cite{PhysRevLett.125.141104}, and x-ray spectroscopy. 
\subsection{The \textsc{HARMPI} code}  
As mentioned in the Introduction, there are several GRMHD codes that are able to construct and evolve accretion disks around BHs. Many of them can be seen in Ref.~\cite{EventHorizonTelescope:2019pcy}, where a direct comparison between them is examined.

The starting point for the code used in this work is \textsc{harmpi} \cite{Tchekhovskoy2019}. This is a public, parallelized, three-dimensional version of \textsc{harm} \cite{Gammie:2003rj,Noble:2005gf}, one of the first GRMHD codes to be developed. It has been chosen for its accessibility, parallelization and recent release date. Apart from it there are a handful of branches from the initial code worth mentioning here. 
A primary derivative
is i\textsc{harm3d} where the equations are solved on a Cartesian grid in arbitrary coordinates and the fluxes are calculated  using the local Lax-Friedrichs (LLF) method (Ref. \cite{Rus61}).   
A recent version is \textsc{h-arm}, extensively updated with improved performance as the main goal, and some new features~\cite{Liska_2020,Chatterjee_2019}. Another widely used version is \textsc{harm}-Noble, or \textsc{harm3d}~\cite{Noble:2008tm,Noble_2011}. It is a variant similar to  i\textsc{harm} with several modest changes, like the incorporation of agnostic coordinates and spacetime choices, making it in a way generally covariant. Lastly, there is r\textsc{harm} (Ref.~\cite{Qian:2016lyn}) with two new physical parameters that have been added to the system, electric field variable and resistivity (or magnetic diffusivity), serving the purpose of smoother, more organic jets evolution.

In its core, \textsc{harm} solves hyperbolic partial differential equations in conservative form using high-resolution, shock-capturing techniques.
It has been configured with the purpose of solving the relativistic magnetohydrodynamic equations of motion in a stationary black hole spacetime in Kerr-Schild coordinates, with the goal to evolve an accreting disk.
These are (as in e.g. Ref. \cite{Noble:2008tm}) the continuity equation 
%Following is an explicit statement of the equations governing the model.   
%The code has to solve the following equations, namely the continuity equation, 
\begin{equation}
    \nabla_\mu(\rho u^\mu)=0, 
\end{equation}
the equation of local energy conservation
\begin{equation}
    \nabla_\mu T^\mu _\nu = 0, 
\end{equation}
and Maxwell's equations
\begin{equation}
    \nabla_\nu F^{\star \mu \nu}=0,
\end{equation}
\begin{equation}
    \nabla_\nu F^{\mu \nu}=J^\mu .
\end{equation}
Where, $\rho$ is the rest-mass density, $u^\mu$ is the 4-velocity of the fluid, $F^{\mu \nu}$ is the Faraday tensor times $1/\sqrt{4\pi}$, $F^{\star \mu \nu}$ is the dual of this tensor or the Maxwell tensor times $1/\sqrt{4\pi}$, and $J^\mu$ is the 4-current.  
In a flux conservative form the equations can be expressed as 
\begin{equation}
    \partial_t \mathcal{U}(P)=-\partial_i \mathcal{F}^i(P) +\mathcal{S}(P),
\end{equation}
where $\mathcal{U}$ is a vector of ``conserved" variables, $F^i$ are the fluxes, and $S$ is a vector of source terms. Explicitly these are 
\be
   \mathcal{U}(P)&=&\sqrt{-g}[\rho u^t, T^t_t+\rho u^t,T^t_j,B^k]^T\\
    \mathcal{F}^i(P)&=&\sqrt{-g}[\rho u^i, T^i_t+\rho u^i,T^i_j,(b^iu^k-b^ku^i)]^T\\
    \mathcal{S}(P)&=&\sqrt{-g}[0,\,T^\kappa _\lambda \Gamma^\lambda _{t\kappa},\,T^\kappa_\lambda\Gamma^\lambda_{j\kappa},0]^T,
\ee
where $g$ is the determinant of the metric, $\Gamma^\lambda_{\mu \kappa}$ is the metric's affine connection, and $B^i=F^{*it}$
is our magnetic field.

The total stress-energy tensor is the sum of the fluid part,
\begin{equation}
    T^{\mu \nu}_{\textrm{fluid}}=\rho h u^\mu u^\nu +P g^{\mu \nu}, 
\end{equation}
and the electromagnetic part 
\begin{equation}
\begin{split}
    T^{\mu \nu}_{\textrm{EM}}&= F^{\mu \lambda}F^\nu_\lambda-\frac{1}{4}g^{\mu \nu}F^{\kappa \lambda}F_{\lambda \kappa} \\
    &= |b|^2 u^\mu u^\nu + \frac{1}{2}|b|^2g^{\mu \nu}-b^\mu b^\nu,
    \end{split}
\end{equation}
where $g_{\mu \mu}$ is the metric, $h=(1+\epsilon+P/\rho)$ is the specific enthalpy, $P$ is the pressure, $\epsilon$ is the specific internal energy density (note the difference with the deformation parameter $\epsilon_3$), $b^\mu = F^{*\nu \mu}u_\nu$ is the magnetic field 4-vector, and $|b|^2 \equiv b^\mu b_\mu$ is twice the magnetic pressure.

The equations of motion are closed by an equation of state, $P=(\Gamma-1)\rho \epsilon$ or $P=K\rho^\gamma$, where $\Gamma$ is the adiabatic index, and $K$ is the entropy parameter. 

\section{Model\label{sec:model}}
In order to model a thin disk around a Johannsen BH, we had to implement three major extensions (or alterations) to the initial package of \textsc{harmpi}. We had to add the metric functions describing the Johannsen metric, we had to implement a cooling function, so that we can dictate the geometric thickness of the disk and we had to change the initial disk profile, so that it matches the literature that we followed. 

The change of metric is fairly straight forward. We defined globally the parameters $\epsilon_3,\, \alpha_{13} ,\, \alpha_{22},\, \alpha_{52} $ and we wrote the functions to calculate the metric components in both BL and KS coordinates. Then, we interchanged all previous metric functions with our own. When all parameters are zero, the metric gets simplified to the Kerr metric. With this modification, \textsc{harmpi} operates on a deformed spacetime.    

\subsection{Cooling function} \label{Cool_func}
As discussed before, \textsc{harmpi} solves equations in a flux conservative form, meaning that the energy dissipated by the turbulent effects of the fluid remains in the fluid in the form of heat. Thus, without the introduction of a loss term in the energy equation, the disk would constantly get hotter and thicker. Eventually the thermal energy would get captured by the hole or carried away by winds. Since we are interested in the thin disk case, it is essential to release this heat. 

To accomplish efficient cooling, we implement a cooling function, following the work of Ref.~\cite{Noble:2008tm}. First, we define vertical thickness. There are a few ways to calculate it; in this work we use the vertical density moment, as it does not need the assumption of a Gaussian profile: \begin{equation}\label{eq:thick}
    H(r)=\frac{\int d\theta d\phi \sqrt{-g}\sqrt{g_{\theta \theta}}\rho |\theta - \pi/2|}{\int d\theta d\phi \sqrt{-g}\rho}\,.
\end{equation}
Often, we will use the variable $\tilde{H}=H/r$ instead of $H$. The temperature of the disk that should produce a desired aspect ratio $H/r$ in Newtonian gravity is
\begin{equation}
   T_* =\frac{\pi}{2}\left[\frac{H}{r}r\Omega(r)\right]^2 .
\label{eq:temp}\end{equation}
For the calculation of $\Omega(r)$ we deviate from Ref.  \cite{Noble:2008tm}. Instead of using the relativistic Keplerian frequency $\Omega(r)=1/(r^{3/2}+a/M)$, we calculate $\Omega$ using the geodesics of a circular orbit on the equatorial plane~\cite{Bambi:2016sac}
\begin{equation}
    \Omega(r)=\frac{-\partial_r g_{t\phi}+\sqrt{(\partial_r g_{t\phi})^2-\partial_r g_{\phi \phi}\partial_r g_{tt}}}{\partial_r g_{\phi\phi}}.
\end{equation}
Since we have a more generic metric it seems very good choice to use a more generic expression for $\Omega(r)$. That way when we use a non-Kerr metric, the changes in the metric components are embedded in the cooling function. For radii smaller than ISCO, we use $\Omega(r_{\rm ISCO})$ in Eq.~\ref{eq:temp}. 

The cooling function $\mathcal{L}$, which is basically the rate at which energy is radiated per unit proper time in the fluid frame, is defined as
\begin{equation}
    \mathcal{L} = s\Omega \rho \epsilon [Y-1+|Y-1|]^q, 
\end{equation}
where $Y=(\Gamma-1)\epsilon/T_*$,  $s=1$ is a factor of proportionality and $q=0.5$ dictates how fast the cooling acts when the fluid is above the desired temperature ($T_*$). Note how the absolute value of ($Y-1$) acts as a switch, turning off cooling whenever the fluid's temperature is below $T_*$. The free parameters were set to their default values, following Ref.~\cite{Noble:2008tm}.

To express the radiation we assume that it is isotropic in the fluid's frame, and so $F_\nu$, the amount of radiated energy-momentum per unit 4-volume in the coordinate frame is 
\begin{equation}
    F_{\nu}= \mathcal{L} u_{\nu}\,.
\end{equation}
Finally, we use their assumption, that the radiation described by this loss term acts in all the volume of the disk, and so the local energy conservation equation reads 
\begin{equation}
    \nabla_\mu T^{\mu}_{\nu} = -F_{\nu}\,.
\end{equation} 
With this equation the procedure is complete and can be summarized as follows. 
%A brief overview of the technique follows.
First, we set the desired aspect $H/r$ and we calculate the temperature $T_*$ that would keep a disk at this thickness. We construct the cooling function $\mathcal{L}$, and we implant it in the local energy conservation equation.
%\subsection{Johannsen metric}
%The approach to this extension is fairly simple, and straight forward. We needed to provide the code with the Johannsen metric in both BL and KS coordinates. We defined globally the parameters $\epsilon_3,\, \alpha_{13} ,\, \alpha_{22},\, \alpha_{52} $, and then we wrote a few functions to calculate these components. When all parameters are zero, the metric gets simplified to the Kerr metric. Otherwise the code creates a deformed space time. \ar{This is probably unnecessary?}   

%\section{Simulation\label{sec:simulation}}
%In order to set up a particular simulation, numerous parameters have to be defined. In this section, we will first go through the creation of the initial disk profile, and then through the specification of all the important parameters that went into our simulation.   
\subsection{Initial disk profile\label{sec:initialdisk}}
In principle, the initial disk profile should be irrelevant for the final thin disk profile we are simulating. Still, we change the default initial disk profile in the original version of \textsc{harmpi}, taken from Ref.~\cite{1976ApJ...207..962F}, to the one used more often in literature (e.g., in Refs.~\cite{De_Villiers_2003, Shafee:2008mm, Noble:2008tm,2010ApJ...711..959N, Noble_2011, 2013ApJ...769..156S, Schnittman_2016, 2016ApJ...826...52K, Kinch_2019}), given in Ref.~\cite{1985ApJ...288....1C}. The first benefit from this is that the initial disk is a bit thinner. More importantly, it is easier for us to compare our results with the literature. %Following is the definition of the disk model and the way we implemented it in the code. 
We now discuss this initial disk profile, following the notation of Ref.~\cite{De_Villiers_2003}. 
%The disk solution is described in a more comprehensive way in  then Chakrabarti (1985) \cite{1985ApJ...288....1C} and we follow a similar route here. 

The starting point of this idea is a solution where the angular velocity in the disk has a power law form
\begin{equation}
    \Omega = \eta \lambda^{-q} \,,
\end{equation}
where $\eta$ is a constant, $q$ is a positive parameter, and $\lambda$ is given by
\begin{equation}
    \lambda^2= \frac{l}{\Omega}=l\frac{g^{tt}-lg^{t\phi}}{g^{t\phi}-lg^{\phi\phi}} \,.
\end{equation}
 It is easy to show that $\lambda$ is the cylindrical radius in Newtonian gravity, i.e. $\lambda=r\sin\theta$. Note, that $l$ and $\Omega$ are defined as \begin{equation}
    l = -\frac{u_\phi}{u_t} \, ; \, \Omega = \frac{u^\phi}{u^t} \, . 
\end{equation} 
We start with the equation of momentum evolution in the hydrodynamic limit, with the assumptions of time independence, axisymmetry and no poloidal motion,

\begin{equation}
    \frac{\partial_j(P)}{\rho h}= -\frac{u^2_t}{2} \partial_j(u^{-2}_t)+u^2_t\left(-\partial_j g^{t \phi} +l\partial_j g^{t\phi}\right)\partial_j l\, ,
\end{equation}
where $P$ is the pressure, $h$ is the specific enthalpy and $u^{-2}_t = g^{tt} - 2lg^{t\phi} +l^2 g^{\phi \phi}$, a result which follows from $u_\mu u^\mu =-1$, considering $u_r = u_\theta = 0$. 

Furthermore, by imposing constant entropy, i.e., $TdS=0$, and using $dh=dp/\rho$, it follows that  
\begin{equation}
    \frac{\partial_jh}{h}= -\frac{1}{2} \frac{\partial_j(u^{-2}_t)}{u^{-2}_t}+u^2_t\left(-\partial_j g^{t \phi} +l\partial_j g^{t\phi}\right)\partial_j l\, .
\end{equation}
The next step is to integrate this equation, and for simplicity the assumption $\Omega \equiv \Omega(l)$ is used, along with the relation $\Omega=(g^{t\phi} - lg^{\phi \phi})/(g^{tt}-lg^{t\phi})$. The expressions for $l$ and $\Omega$ can be rewritten as   
\begin{equation}
    l=\eta \lambda^{2-q}\, ; \, \Omega =\eta^{-2/(q-2)} l^{q/(q-2)} \equiv kl^\alpha,  
\end{equation}
where $\alpha = q/(q-2)$. 
A general solution arises when setting enthalpy at the inner edge of the disk to zero ($h_{ \rm in} =0$) and determining the surface binding energy $u_{ \rm in} $
\begin{equation}
    h(r,\theta)=\frac{u_{ \rm in} f(l_{ \rm in} )}{u_t(r,\theta)f(l(r,\theta))}\,,
\end{equation}
where 
\begin{equation}
    f(l)=|1-kl^{(a+1)}|^{1/(a+1)}\, .
\end{equation}
Using the equation of state and the definition of enthalpy, the internal energy of the disk is 
\begin{equation}
    \epsilon(r,\theta) = \frac{1}{\Gamma}\left\{\frac{u_{ \rm in} f(l_{ \rm in} )}{u_t(r,\theta)f(l(r,\theta))}-1\right\}\,,
\end{equation}
and the density  
\begin{equation*}
    \rho=[\epsilon(\Gamma-1)/\kappa]^{\sfrac{1}{(\Gamma-1)}} .
\end{equation*} 

The surface binding energy $u_{ \rm in} $ can be explicitly defined if we provide $l_{ \rm in} $. Lastly, we provide a way to calculate   $\eta$ (or $k$). We make use of  

\begin{equation}
    \frac{l}{l_{ \rm in} }=\left( \frac{\lambda}{\lambda_{ \rm in} } \right)^{2-q}\,,
\end{equation}
so it follows that 
\begin{equation}
    \eta=\frac{l_{ \rm in} }{(\lambda_{ \rm in} )^{2-q}}\, ; \, k = \eta^{-2/q-2}\,.
\end{equation}
 Eventually, we can calculate enthalpy ($h$) at any point of the disk, and subsequently $\rho$, as well as the components $u_t$, $u_\phi$ of the 4-velocity, which gives us the complete profile of the disk.

 \begin{figure}[!htb]
\begin{subfigure}{0.4\textwidth}
\includegraphics[width=1\linewidth]{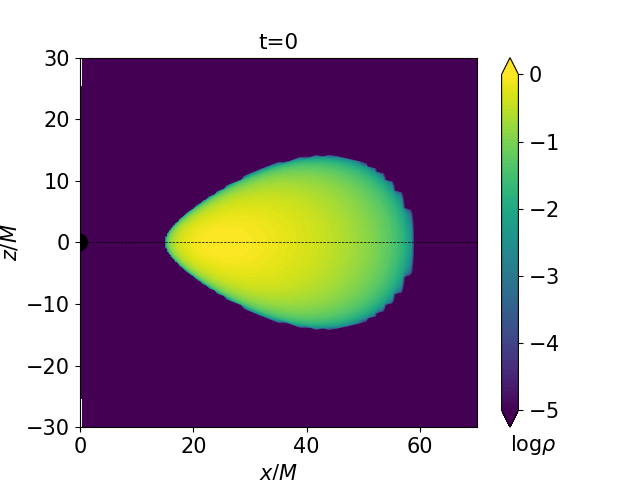} 
\end{subfigure}
\caption{A snapshot of the initial disk profile from a Kerr BH with $a=0.9$. The density is in code units, normalized to 1. The black dashed line denotes the equatorial plane.}
\label{iniprof}
\end{figure}

\section{Simulation parameters\label{sec:simulation}}
%\subsection{Description of models}

In our effort to explore the effect that deformation parameters bring to the models as broadly as possible, we use various combinations of parameters. Some of our simulations are in pure Kerr background, which serve as benchmark.

The simulations reported here are 2D that use a  $192\times 192$ grid in the radial and polar  directions, with $r \in[ \sim R_{\rm hor},120]$, and $\theta \in [0.05\pi,0.95\pi]$. To check for convergence we run simulations both with higher and lower resolutions and the chosen configuration was the optimal combination of resolution and computational time. 
%was the best  with $384\times 192$, as well as $762\times 192$, and we saw that it does not affect the quality of the results much, so we stick to the cheapest (computationally) case. 
We also run a 3D model with resolution $96\times 96\times 64$, but it was too computationally  expensive and long for the number of simulations that we performed. The details of the 3D case are presented in Appendix \ref{App1}, and an extension of the present work by using 3D simulations is planned. 

We set the the inner edge of the initial disk to $R_{ \rm in} =15M$ and the parameter $q=1.68$. The specific angular momentum $l_{ \rm in} $ was adjusted differently in every case (exact values given in Tab.~\ref{cases}) to result always in a similar disk with the same pressure maximum point $R_{\rm pmax}=25$. Fig.~\ref{iniprof} shows the initial disk profile, identical to the solution of Ref.~\cite{1985ApJ...288....1C} and same for all cases. 

The adiabatic index, and the entropy parameter are set to $\Gamma =5/3$ and $K =0.01$. The parameter $\beta$, defined as the ratio of the gas pressure to the magnetic pressure, using the volume-integrated gas pressure divided by the volume-integrated magnetic energy density in the initial torus, was set equal to $100$. The desired thickness was set to $\tilde{H}_d=(H/r)_{\rm desired}=0.12$. 
We control the discretization of the grid through the parameter $h_{\textrm{KMS}}=0.3$, that produces smaller cells as the radius gets smaller and the closer we are at the equatorial plane. The floor for the background density has been set at $\rho_{\textrm{floor}}= 10^{-7}$. Turbulence was seeded by adding random perturbations to $u$ at the $1\%$ level. All simulations were evolved for $t=[0,15000M]$, and we recorded the state of the disk every $t=20M$.

Tab.~\ref{cases} lists a subset of simulations we performed and those we will discuss in the rest of the paper. Our choices are driven by the anticipation, based on previous analyses of x-ray reflection spectroscopy, that there exists a degeneracy between BH spin and deformation parameters based on the size of ISCO. Thus, ISCO provides a measure of the BH field strength and a rubric for constructing the simulation set.  
%The motivation for the choice of these sets of parameters, is that we want to explore differences between cases with comparable spin or ISCO. Since ISCO proved to be a more dominant parameter 
We simulate cases with moderate ISCO radii (cases I, II, III) representing a BH neighborhood with moderately strong fields, and cases with very small ISCO radii (cases IV, V, VI) representing neighborhoods with extremely strong gravitational fields. From now, the simulations will be referred to using the case number from Tab.~\ref{cases}.    
\begin{table}[h]
    \caption{ List of all the simulations discussed in the text. Deformation parameters when not mentioned are set to zero. Spin and ISCO are in given in units of $M$ and the values for mass are given in code units.\label{cases}}
    \centering
    \begin{tabular}{c|c|c|c|c|c}
    \hline
        \textbf{Case}& \textbf{Spin}& \textbf{Non-Kerr Parameter}&\textbf{ISCO}& \textbf{$l_\text{in}$}&\textbf{M}  \\
        \hline
        I&0.9&0&2.32&4.58&30500\\
        II&0.678&$a_{13}=-2$&2.32&4.61&35000\\
        III&0.9&$a_{52}=3$&2.32&4.58&30500\\
        IV&0.976&0&1.66&4.58&35200\\
        V&0.9&$a_{13}=-1.5$&1.66&4.58&30500\\
        VI&0.998&0&1.237&4.576&34000\\
        
        IX$_{\rm 3D}$&0.9&0&2.32&4.58&18600\\
        
        \hline
    \end{tabular}
\end{table}

%For every one of the simulations listed, we create graphs and present information regarding the structure, the stability and the radiation of the disk. It is redundant to exhibit all graphs from all simulations, so we choose to show the most important and interesting from every case. %We also make comparisons in cases which we think are comparable.

\section{Results\label{sec:results}}

\begin{figure*}[!htb]
\begin{subfigure}{0.3\textwidth}
\includegraphics[width=1\linewidth]{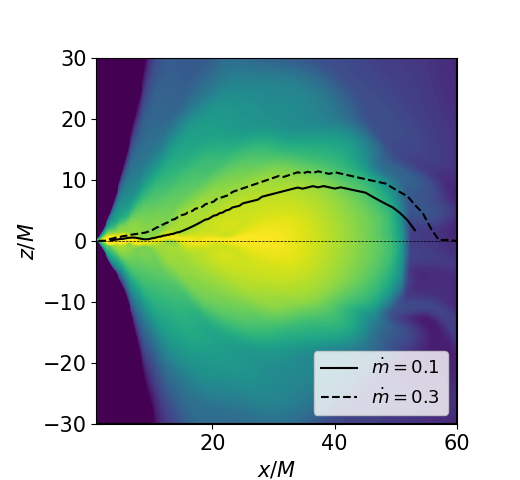} 
\caption{Case I}
\end{subfigure}
\begin{subfigure}{0.3\textwidth}
\includegraphics[width=1\linewidth]{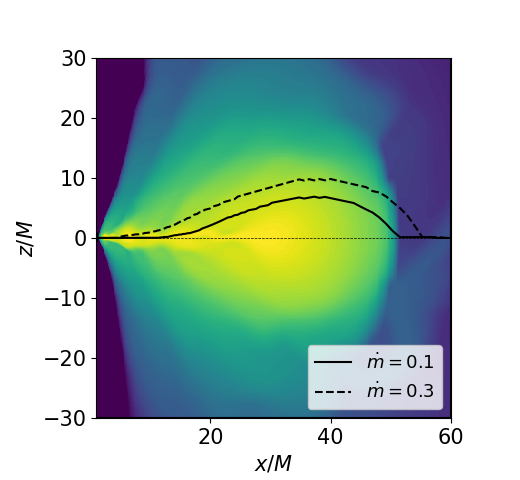}
\caption{Case II}
\end{subfigure} 
\begin{subfigure}{0.3\textwidth}
\includegraphics[width=1\linewidth]{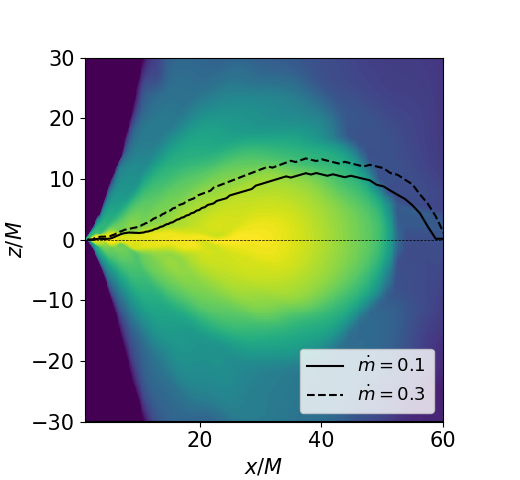} 
\caption{Case III}
\end{subfigure} \\
\begin{subfigure}{0.3\textwidth}
\includegraphics[width=1\linewidth]{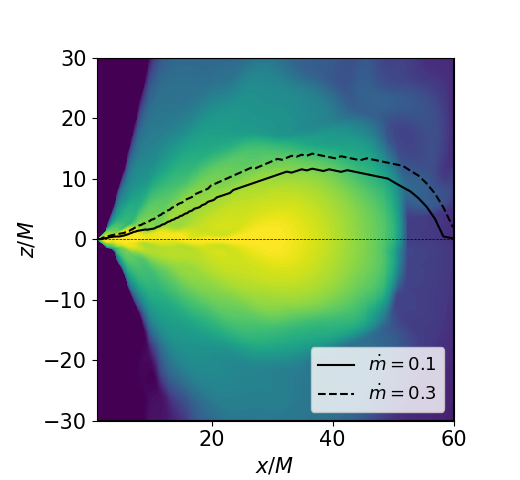} 
\caption{Case IV}
\end{subfigure}
\begin{subfigure}{0.3\textwidth}
\includegraphics[width=1\linewidth]{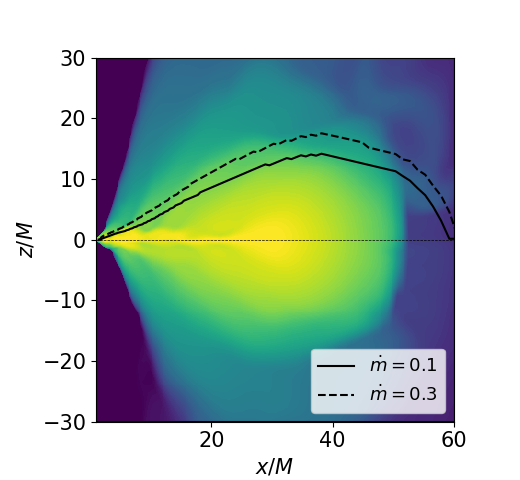} 
\caption{Case V}
\end{subfigure}
\begin{subfigure}{0.3\textwidth}
\includegraphics[width=1\linewidth]{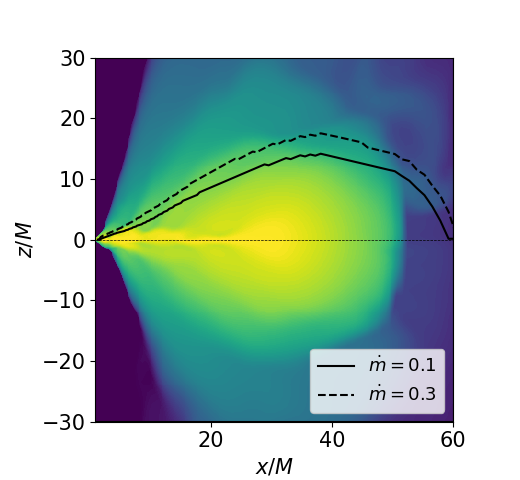} 
\caption{Case VI}
\end{subfigure}
\caption{Snapshots of the disk at the end of the simulation ($t=15000M$) for different cases. The color coding remains the same as in Fig.~\ref{iniprof}. The solid and dashed black lines mark the time averaged height $(\tilde{H}_{\rm ph})$ of the surface of constant optical depth $\tau=1$ for two indicative values of $\dot{m}$, and the dotted black line marks the equator. See text for more details.\label{eveprof}} 
\end{figure*}
\subsection{Disk properties\label{subsec:diskprop}}

In this first section of our results, we examine some fundamental properties of GRMHD disks regarding their structure and stability. Ensuring that the disks pass these first tests is essential for the credibility of the more complex calculations to follow.  

In Fig.~\ref{eveprof} we exhibit the disk at the last time frame from every simulation (the black over-plotted lines are discussed in a later section). This is a typical image of a stable thin accretion disk in stability. The disks reach stability around $t=6000-8000M$ and remain stable till the end of the simulation. This is why all averaging is taking place in the interval $t=8000-15000M$. 
%The main body of the disk and the inner torus is clearly distinguishable, with intense yellow in our case. Around it, with light blue we can see the corona enveloping the main disk.  

\begin{figure*}[!htb]
\centering
\begin{subfigure}[t]{0.33\textwidth}
\includegraphics[width=\columnwidth]{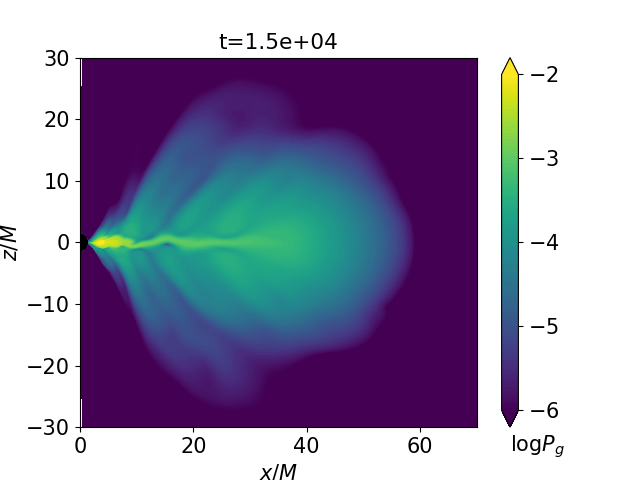}
\caption{Thermal gas pressure}
\end{subfigure}%
\begin{subfigure}[t]{0.33\textwidth}
\includegraphics[width=\columnwidth]{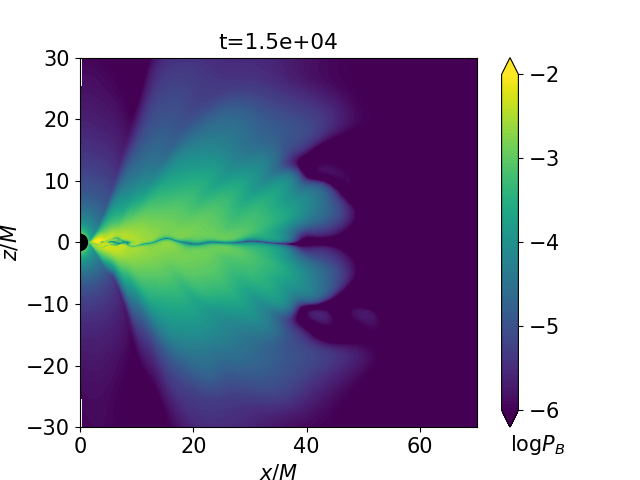} 
\caption{\label{magpre}Magnetic pressure}
\end{subfigure}%
\caption{Snapshots from case I of the magnetic and thermal gas pressure. The images were taken at the end of the simulation ($t=15000M$).}
\label{presprof}
\end{figure*}

Next, in Fig.~\ref{presprof}, we show profiles of the magnetic and gas pressure from case I. We can observe the thermal gas pressure $P_g$ having a similar behavior to the density $\rho$, peaking in the equatorial region. The magnetic pressure $P_B$, on the other hand, is practically zero in this region and reaches its maximum in an area very close to the hole (where the density of magnetic lines is larger), where both the gas and magnetic pressures are high. 
%\begin{widetext}

\begin{figure*}[!htb]
\begin{subfigure}{0.33\textwidth}
\begin{center}
    \includegraphics[width=1\linewidth]{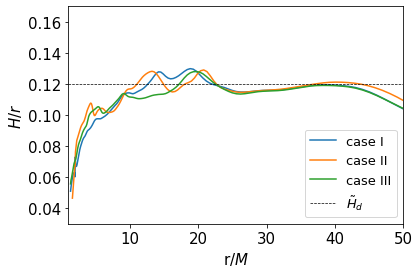} 
    \caption{Cases I, II, III}
    \label{Hor1-2-3-10}
\end{center}   
\end{subfigure}%
\begin{subfigure}{0.33\textwidth}
\begin{center}
    \includegraphics[width=1\linewidth]{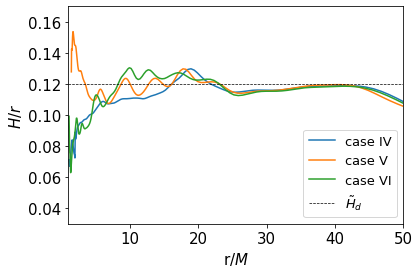}
    \caption{Cases IV, V, VI}
    \label{Hor4,5,7}
\end{center}
\end{subfigure} 
\begin{center}
\caption{ Vertical thickness $H/r$, averaged over $t=8000M-15000M$, with respect to $r$. $\tilde{H}_d$ denotes the intended thickness ($= 0.12$).}
\label{Hor}
\end{center}
\end{figure*}

%\end{widetext}
In Fig.~\ref{Hor} we calculate the vertical thickness $H/r$ as defined in Eq.~\ref{eq:thick}. We conclude that in general, despite the different spins and deformation parameters, all models came very close to the goal ($\tilde{H}_d =0.12$). This result is of great importance, as it shows that the simulated disks indeed grew thinner, and the cooling function worked as intended. In case V there is a unique feature of the thickness rising sharply in the inner region of the disk. %This seems to be an effect of the high spin in combination with the negative deformation parameter $a_{13}=-1.5$. 
We discuss this case further below.    

\begin{figure*}[!htb]
\begin{subfigure}{0.3\textwidth}
\includegraphics[width=1\linewidth]{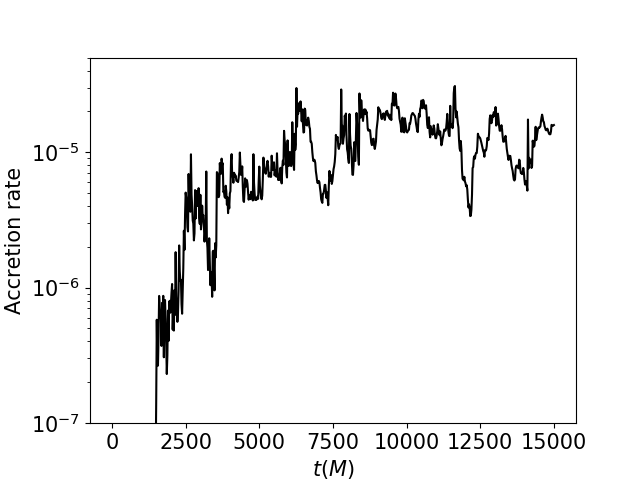} 
\caption{Case I }
\label{Aca1}
\end{subfigure}
\begin{subfigure}{0.3\textwidth}
\includegraphics[width=1\linewidth]{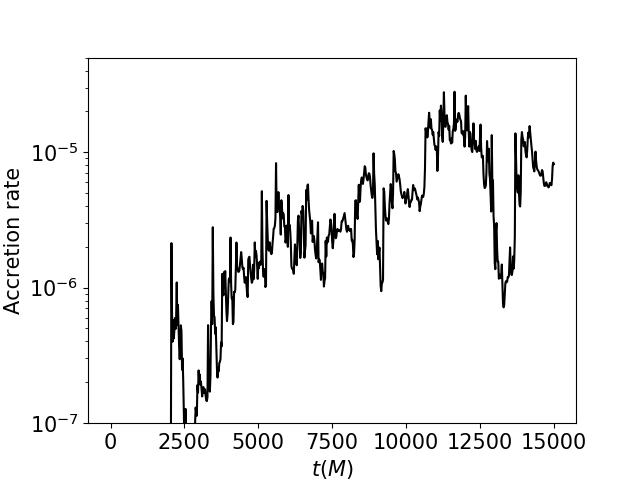} 
\caption{Case II }
\label{Aca2}
\end{subfigure}
\begin{subfigure}{0.3\textwidth}
\includegraphics[width=1\linewidth]{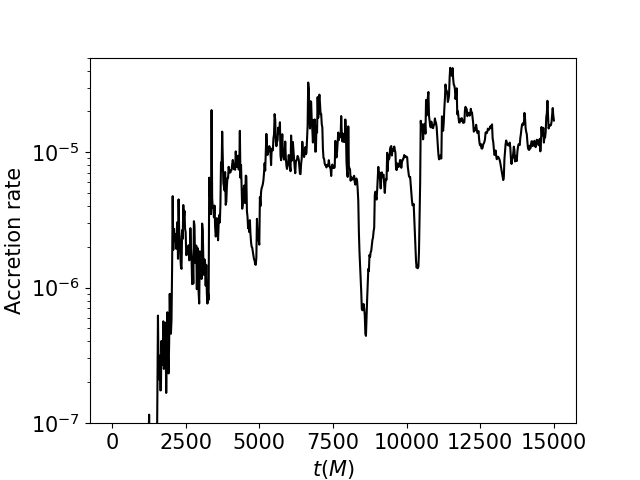}
\caption{Case III }
\end{subfigure} \\
\begin{subfigure}{0.3\textwidth}
\includegraphics[width=1\linewidth]{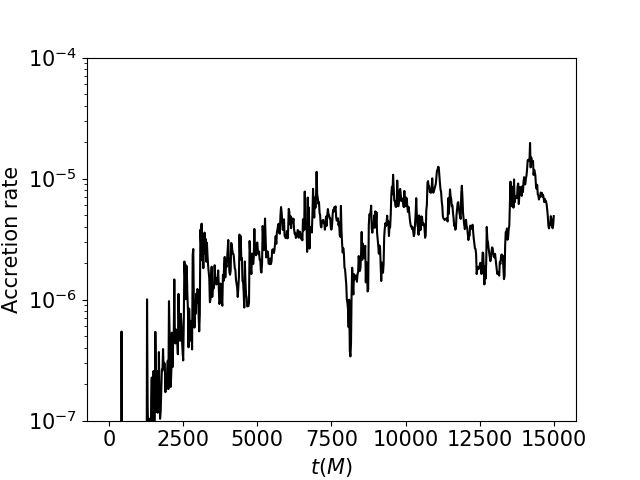} 
\caption{Case IV} 
\end{subfigure}
\begin{subfigure}{0.3\textwidth}
\includegraphics[width=1\linewidth]{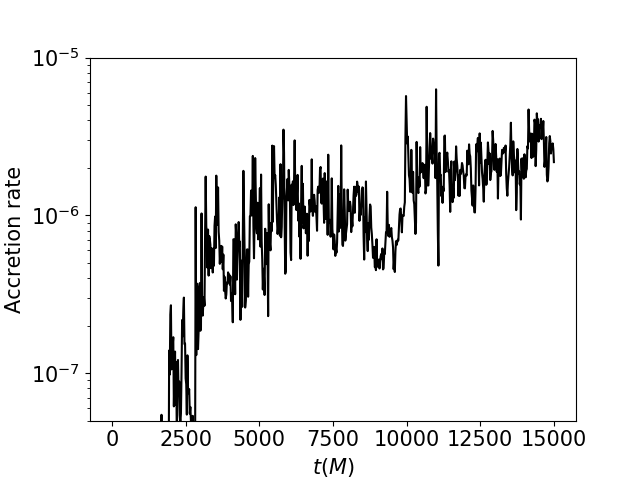} 
\caption{Case V} 
\label{Aca5}
\end{subfigure}
\begin{subfigure}{0.3\textwidth}
\includegraphics[width=1\linewidth]{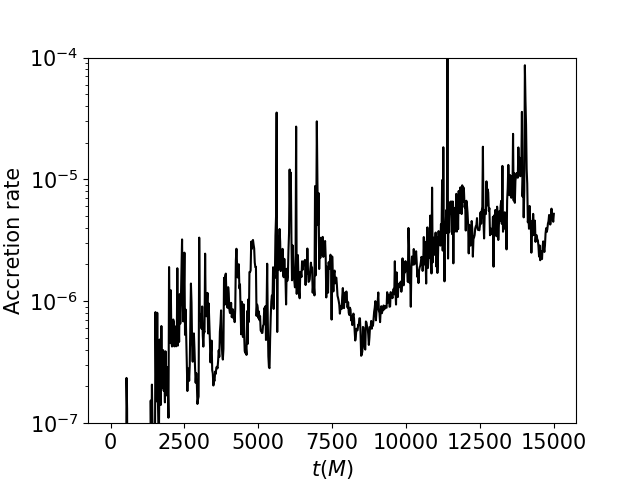} 
\caption{Case VI }
\end{subfigure}
\caption{ Accretion rate at the horizon, in units of $M_{\rm code}/M$, as a function of time for different cases.} 
\label{Aca}
\end{figure*}
The next graphs are devoted to the stability of the disks. We test this by examining the behavior of the accretion rate both with respect to radius and time. We claim that the disks are in a steady state after $t=8000M$, meaning that we expect a stable accretion. For the accretion as a function of time the fluctuations are unavoidable with any numerical approach, since the sampling is far from smooth. In Fig.~\ref{Aca}, we present the accretion rate at the horizon, $\dot{M}(r_{\rm hor})$, with respect to time. We can see that the accretion rate has stabilized and remains quite low. These graphs are similar to what is seen in typical GRMHD simulations (see, for instance, Fig. 2 from \cite{De_Villiers_2003,2013ApJ...769..156S,king2016} as well as Fig. 13 from \cite{Porth:2016rfi}). 

\begin{figure*}[!htb]
\begin{center}
\begin{subfigure}{0.33\textwidth}
\includegraphics[width=1\linewidth]{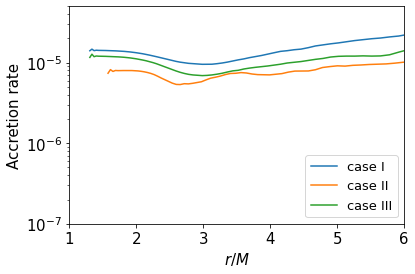}
\caption{Cases I, II, III}
\label{Acb1}
\end{subfigure}
\begin{subfigure}{0.33\textwidth}
\includegraphics[width=1\linewidth]{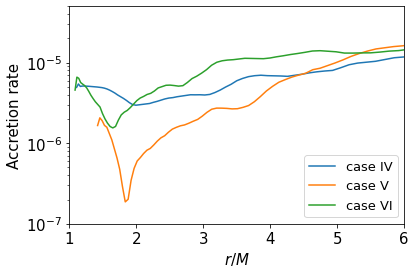}
\caption{Cases IV, V, VI}
\label{Acb2}
\end{subfigure}
\caption{ Time averaged accretion rate, in units of $M_{\rm code}/M$ and averaged over $t=8000M-15000M$, with respect to $r$.}
\label{Acb}
\end{center}
\end{figure*}

In Fig.~\ref{Acb}, the graphs of accretion (averaged in time) with respect to $r$ are shown. The accretion rate is calculated as
\begin{equation}
    \dot{M}(r) =\int_{2H} \int_\phi \rho(r,\theta,\phi) u^r(r,\theta,\phi) g_{\text{det}}(r,\theta,\phi)/M_{\text{code}},
\end{equation}
where $M_{\text{code}}$ is the mass of the disk in code units. For cases I-IV and VI, the rate stays in a close range to the value at the horizon. In case V though, there is a significant drop close to the ISCO. In fact, the drop in accretion rate seen here happens at similar radii where we saw a rise in thickness in Fig.~\ref{Hor}. One possible explanation for this feature is the dampening of magnetorotational instabilties (MRIs) that can lead to a halt in accretion, which can occur especially in 2D simulations. Another possibility is the extreme nature of the parameter choice in this case: $a_* = 0.9$ and $\alpha_{13} = -1.5$ lies very close to the region of the parameter space where pathologies appear in the spacetime (for instance, naked singularities, see Fig. 6 in Ref.~\cite{Johannsen2015}).

%This is due \sn{can you prove it is due to these reasons?} \ar{Not very robuslty.. It is rather a cumulative impression from the extensive work with the different cases. By that i refer to the troubles I faced in my first efforts to simulate these spacetimes, like, grid, resolutions, and skyrocketing values close to ISCO sometimes (if you recall it).  It's even hard to try and find a way to prove that without running a few more sims, to test and compare.} to the fact that both these cases are rather extreme, for different reasons. Case $6$ has a very extreme spin, and the code cannot resolve very well the gas at such a small radius. case V, as pointed before, exhibits in general some special features. The combination of spin and deformation parameter for this case makes the spacetime marginally stable, according to \cite{Johannsen2015}, and we believe that this is what we are observing both in figures \ref{Hor} and \ref{Acb}. Nevertheless, we compensate for these anomalies in accretion rate for all cases in future calculations wherever is needed.

\subsection{Radiative efficiency}
%Fluid frame flux

%Luminosity at infinity

%Integrated radiative efficiency

%\sn{Apart from the thesis, this subsection should also take some inspiration from Noble et al 2011 Sec. 3.}\\

After establishing that the disk evolves as intended in a stable fashion and achieves the desired thickness, we turn our attention to observables. There are different ways to measure the radiation from a GRMHD disk. The first one, and most straight forward, is to consider the radiation produced from the cooling function $\mathcal{L}$, as described in section \ref{Cool_func}. Apart from giving us a way to control the vertical thickness, the function provides a self-consistent way of comparing emission from the simulated disk with that expected in a standard NT model. For doing so, we use the angle-averaged, fluid-frame luminosity per unit area, as in Ref.~\cite{Noble:2008tm}, 
\begin{equation}
     F_{\rm ff}(r) =  \frac{\int \int dx^{(\phi)}dx^{(\theta)}\mathcal{L}}{\int dx^{(\phi)}|_{\theta=\pi/2}},
\end{equation}
where each component of the vector $dx^{(\mu)} = e^{(\mu)}_\nu dx^\nu$ represents the extent of a cell’s dimension as measured in the fluid element’s rest frame, and $e^{(\mu)}_\nu$ is the orthonormal tetrad that transforms vectors in the Boyer-Lindquist coordinate frame to the local fluid frame (see, e.g., Ref.~\cite{Beckwith:2008pu}, for explicit expressions of the tetrad). The vector $dx^\nu$ is the Boyer-Lindquist coordinate frame version of the Kerr-Schild vector $dx^\nu_{\rm KS} = [0,\Delta r,\Delta\theta,\Delta\phi](r, \theta, \phi)$, where $\Delta r, \Delta\theta, \Delta\phi$ are the radial, poloidal, and azimuthal extents of our simulation’s finite volume cell located at $(r, \theta, \phi)$. Eventually, for a 2D simulation the radiated flux per unit area in the fluid frame is given as,
\begin{equation}
   F(r)= \frac{\dot{M}}{\dot{M}(r)} \int \int( e^3_1\Delta r +e^3_2\Delta \theta )(e^2_1\Delta r + e^2_2 \Delta \theta)\mathcal{L}.
\end{equation}
We have inserted the factor $\dot{M}/\dot{M}(r)$ to compensate for any anomalies from the accretion rate.
The flux for the NT model can be calculated as, 
\begin{equation}
    F_{\rm NT}(r) = \frac{\dot{m}}{4\pi g}\frac{\partial_r\Omega}{(E-\Omega L)^2} \int_{r_{\mathrm{isco}}}^r (E-\Omega L)\partial_r L dr,
\label{eq_Fnt}
\end{equation}
where $E$ stands for specific energy, $L$ for specific angular momentum and $\Omega$ for the angular frequency. All quantities are considered in the equatorial plane, and can be calculated as,
\begin{subequations}
\begin{align}
\Omega&=\frac{-\partial_rg_{t\phi} \pm \sqrt{(\partial_rg_{t\phi})^2-(\partial_rg_{tt})(\partial_rg_{\phi \phi})}}{\partial_rg_{\phi \phi}},\\
E &= -\frac{g_{tt}+\Omega_{t\phi}}{\sqrt{-g_{tt}-2\Omega g_{t\phi}-\Omega^2g_{\phi \phi}}},\\
L &= \frac{g_{t\phi}+\Omega g_{\phi \phi}}{\sqrt{-g_{tt}-2\Omega g_{t\phi}-\Omega^2g_{\phi \phi}}}.
\end{align}
\end{subequations}
 Note that when we calculate the $F_{\rm NT}$ to match a case with non zero deformation, we use the deformed metric components to calculate the quantities necessary ($\Omega,\; L,\; E$). The factor $\dot{m}$ in Eq.~\ref{eq_Fnt} is a scaling parameter and takes a value between $0.05$ and $0.2$.

\begin{figure*}[!htb]
\begin{subfigure}{0.3\textwidth}
\includegraphics[width=1\linewidth]{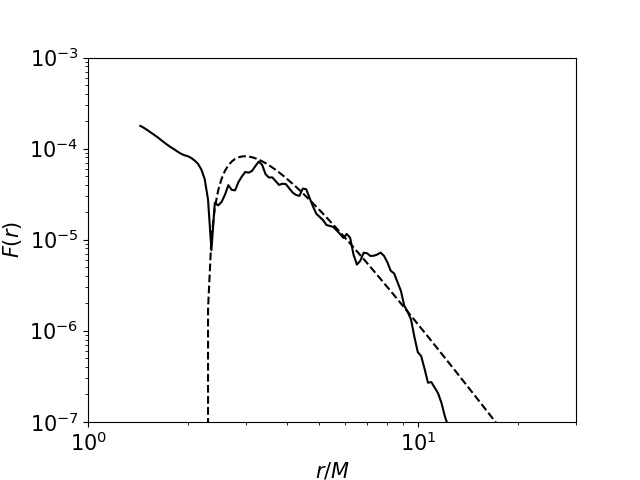} 
\caption{Case I}
\label{Fr1}
\end{subfigure}
\begin{subfigure}{0.3\textwidth}
\includegraphics[width=1\linewidth]{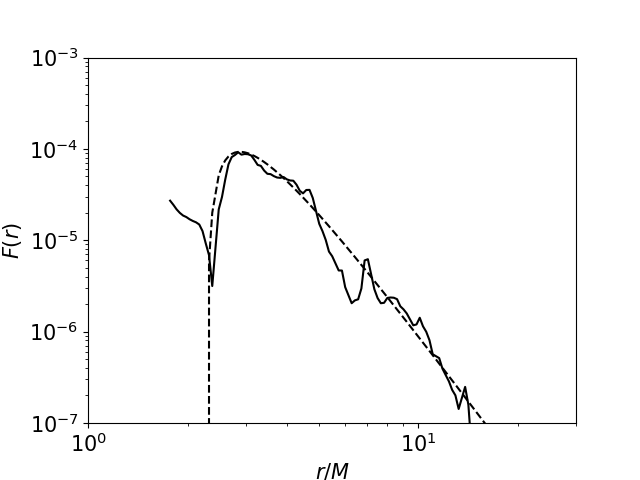}
\caption{Case II}
\end{subfigure} 
\begin{subfigure}{0.3\textwidth}
\includegraphics[width=1\linewidth]
{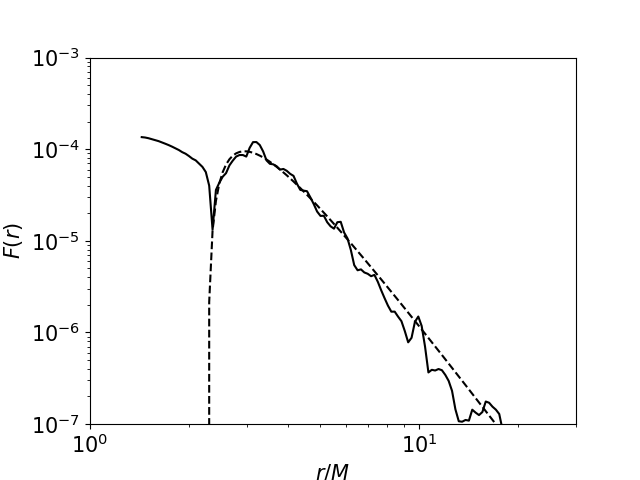} 
\caption{Case III}
\end{subfigure}\\
\begin{subfigure}{0.3\textwidth}
\includegraphics[width=1\linewidth]
{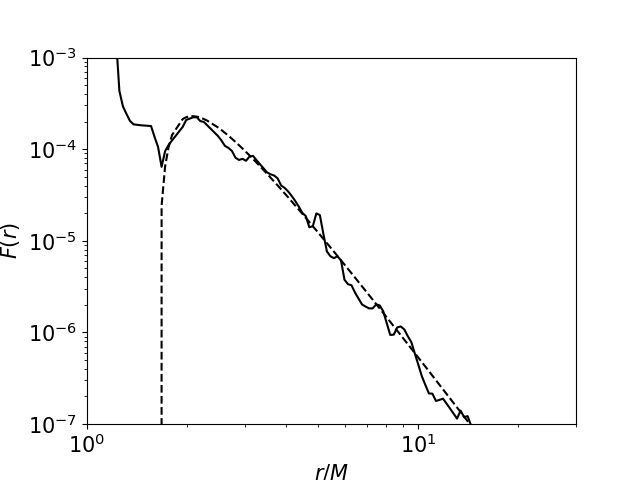} 
\caption{Case IV} 
\end{subfigure}
\begin{subfigure}{0.3\textwidth}
\includegraphics[width=1\linewidth]
{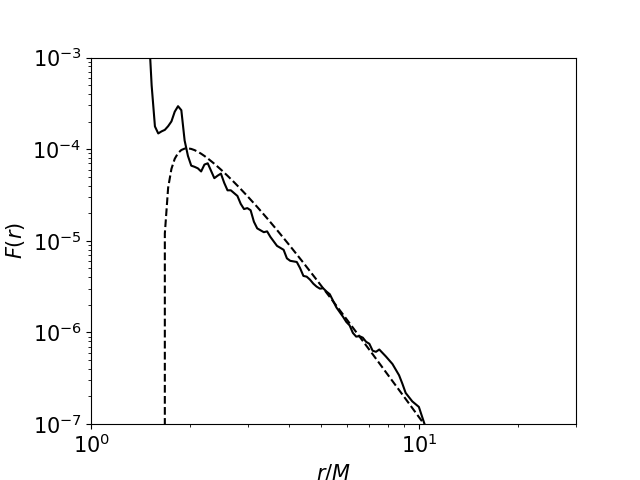} 
\caption{Case V} 
\end{subfigure}
\begin{subfigure}{0.3\textwidth}
\includegraphics[width=1\linewidth]
{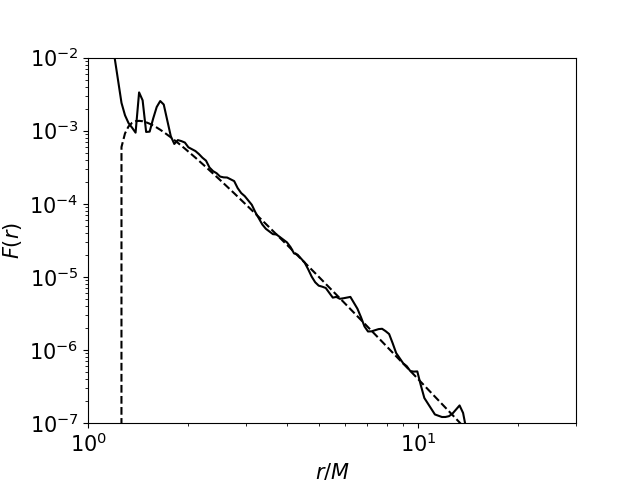} 
\caption{Case VI} 
\end{subfigure}\\
\caption{ Radiated flux per unit area in the fluid frame, averaged over $t=8000M-15000M$, as a function of radius $r$. The dashed line denotes the corresponding line as predicted from the NT model, and described in the text.}
\label{Fr}
\end{figure*}
\begin{figure*}[!htb]
\begin{subfigure}{0.33\textwidth}
\includegraphics[width=1\linewidth]{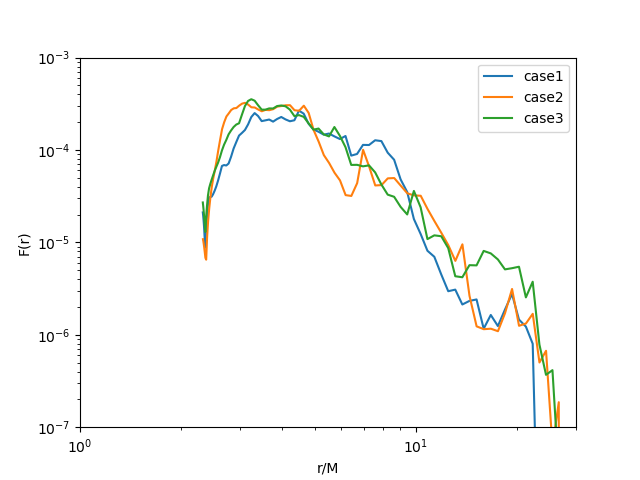} 
\caption{Cases I, II, III}
\end{subfigure}
\begin{subfigure}{0.33\textwidth}
\includegraphics[width=1\linewidth]{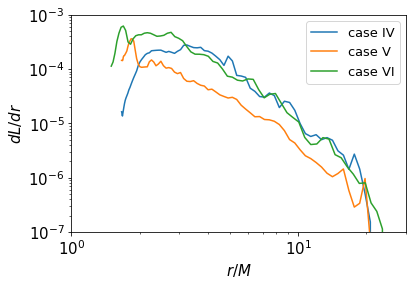} 
\caption{Cases IV, V, VI}
\end{subfigure}

\caption{ Luminosity received at infinity in different cases, based on radiated fluxes of Fig.~\ref{Fr}. See the text for more details.}
\label{dLdr}
\end{figure*}

In Fig.~\ref{Fr} we compare the calculated $F(r)$ from all cases with the predicted flux from the NT model. The flux overall looks in satisfying  agreement with the NT model. This result  is very important as it can vigorously verify the effective performance of the cooling function $\mathcal{L}$ and ultimately the assumption of a NT disk for modelling radiation in general. Here again, we see a deviation from the norm in the inner regions in case V (and to a small extent, case VI). 

In Fig.~\ref{dLdr} we provide the luminosity received at infinity per unit radial coordinate ($dL/dr$), computed following the scheme outlined in Ref.~\cite{Noble:2008tm}, assuming no radiation arrives from inside the ISCO. We again see distinct behavior of case V (and to some extent case VI) compared to all other cases in the inner region, close to the ISCO. Thus, we conclude that the disk structure and radiation in extreme gravity (both pure Kerr with extreme spin and non-Kerr with parameter values near the edge) exhibits interesting features when simulated. We cannot rule out the possibility that these features are due to the 2D nature of the simulations, and assign a deeper investigation to a future work. 

%\subsection{Thermal spectrum}
%This section is a maybe.

\subsection{Reflection spectrum}

We now discuss one of the simplest observables of x-ray reflection spectroscopy: the relativistically broadened iron line. It originates from those thermal photons that interact with, and get up-scattered by, coronal plasma, travel back towards the disk and get reflected. While the complete spectrum includes emission frequencies associated with all the elements (and their ionizations) present in the disk, the K$\alpha$ emission of Iron dominates and for this reason is often the only part of the full reflection spectrum that is detected. For illustration, we therefore choose to present this single emission line. Though at the time of emission it is monochromatic, the motion of the disk, along with the motion of photons in the gravitational field of the BH, broadens it into a spectrum spread across a frequency band. 
\begin{figure*}[!htb]
\begin{subfigure}{0.3\textwidth}
\includegraphics[width=1.\linewidth]{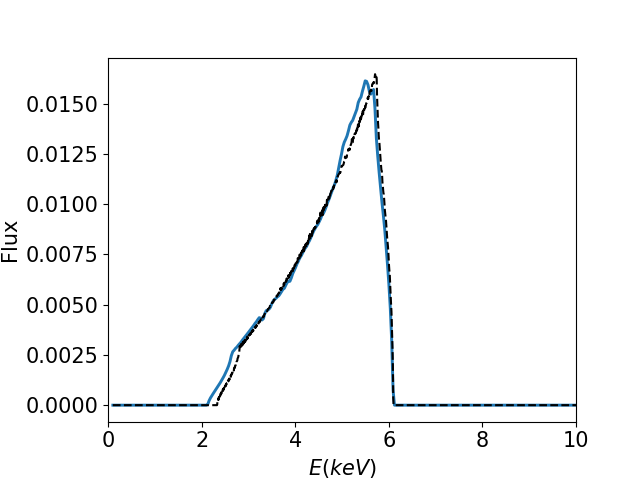}
\caption{Case I}
\label{ironl1}
\end{subfigure}
\begin{subfigure}{0.3\textwidth}
\includegraphics[width=1.\linewidth]{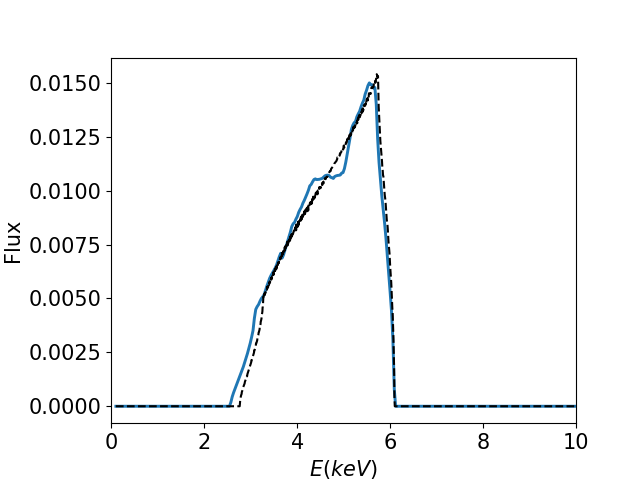}
\caption{Case II}
\end{subfigure}
\begin{subfigure}{0.3\textwidth}
\includegraphics[width=1.\linewidth]{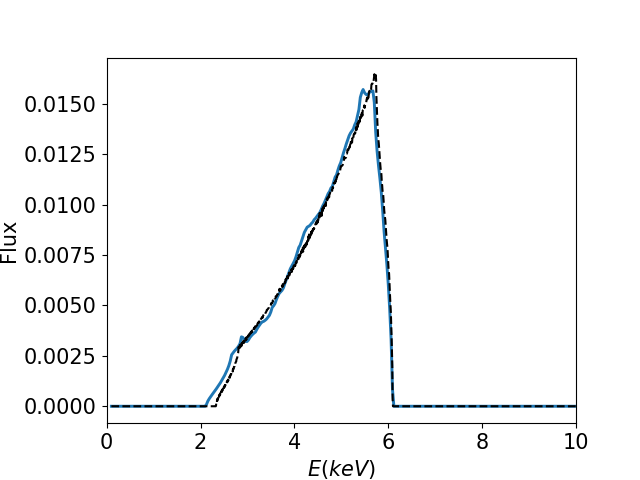}
\caption{Case III}
\end{subfigure}\\
\begin{subfigure}{0.3\textwidth}
\includegraphics[width=1.\linewidth]{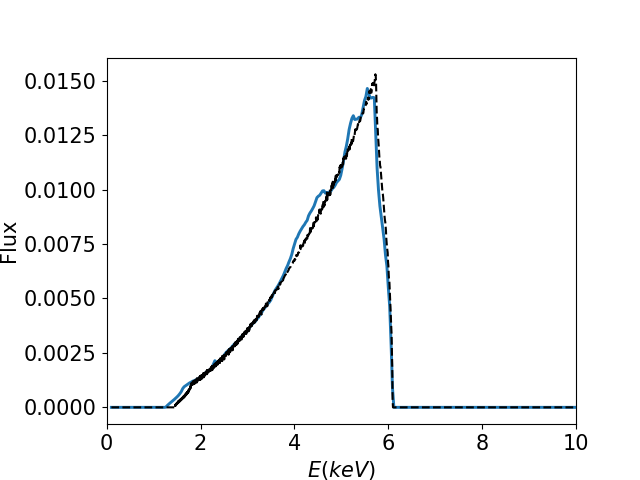}
\caption{Case IV}
\end{subfigure}
\begin{subfigure}{0.3\textwidth}
\includegraphics[width=1.\linewidth]{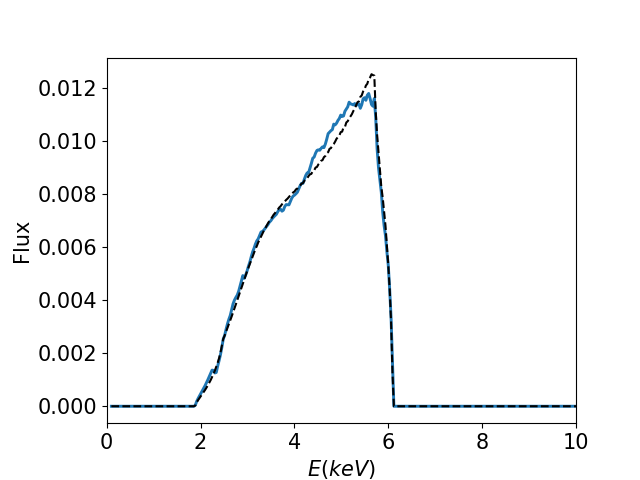}
\caption{Case V}
\end{subfigure}
\begin{subfigure}{0.3\textwidth}
\includegraphics[width=1.\linewidth]{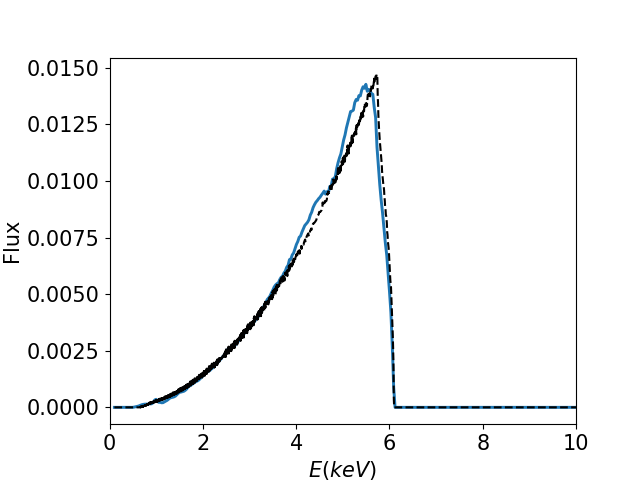}
\caption{Case VI}
\end{subfigure}
\caption{ Iron lines calculated for different cases. with $\dot{m}=0.1$. The blue solid line is from simulations, and the black dotted line from the analytical Novikov-Thorne disk. See the text for more details.} 
\label{csall_IL}
\end{figure*}

%As discussed in the introduction, the iron line is produced after radiation created in the corona, gets reflected on the photosphere. So, 
The first step in order to calculate the Fe K$\alpha$ spectrum is to determine the reflecting surface, defined as $\tau=1$, where $\tau$ is the optical depth. (We follow the notation of Refs.~\cite{2013ApJ...769..156S,2016ApJ...826...52K} for the calculation of the photosphere.)
 %photospheric surface on our disk, compute the flux from that surface and then, using ray-tracing techniques get the reflected part at infinity.
%Following Schnittman et al.(2013)\cite{2013ApJ...769..156S} and Kinch et al. (2016)\cite{2016ApJ...826...52K} we consider the photospheric surface as the surface at which the optical depth, $\tau$=1. 
The optical depth can be calculated as
\begin{equation}
    d\tau = \kappa \rho_{\text{cgs}}r d\theta,
\end{equation}
where $\kappa$ is the Thomson scattering opacity and $\rho_{\textrm{cgs}}$ is the disk density in physical units, which can be calculated by relating it to the code density $\rho_{\textrm{code}}$ as
\begin{equation}
    \rho_{\text{cgs}}= \rho_{\text{code}}\frac{4\pi c^2}{\kappa G M}\frac{\dot{m}/\eta}{\dot{M}_{\text{code}}},
    \label{eq:hph1}
\end{equation}
where $\kappa = 0.4\; {\rm cm}^2 {\rm g}^{-1}$ is the electron scattering opacity, $\dot{m}$ is the Eddington-scaled accretion rate assuming a radiative efficiency $\eta = 0.06$ and $M$ has been set to $10$ \(M_\odot\). The photosphere is given by the following integral: 
\begin{equation}
    \int_{\theta=0}^{\Theta_{\rm top}} d\tau =1, 
    \label{eq:hph2}
\end{equation}
and the height of the photosphere is $H_{\text{ph}}=r\cos{\Theta_{\text{top}}}$. Note that $\theta$ starts from the $z$-axis, in essense $\theta \in(0,\pi). $
In Fig.~\ref{eveprof} we show the photosphere for two values of $\dot{m}$, 0.1 and 0.3, for each simulation case. As expected, the height of the photosphere increases with the rate of Eddington-scaled accretion ($\dot{m}$), since higher accretion rate implies higher densities, and higher density reduces optical depth. The second variable that is able to make a difference in the height of the photosphere (eq. \ref{eq:hph1}, \ref{eq:hph2}) is the code's accretion rate ($\dot{M}_{\text{code}}$). It is visible from the graphs that indeed cases with smaller accretion rates (namely, cases IV, V and VI, see Sec.~\ref{subsec:diskprop}) exhibit higher photospheres for both values of $\dot{m}$. 
%One should remember that the photosphere should not be too large, because further away from the equatorial plane, the 4-velocities to calculate the iron lines, are getting more varying (\sn{I do not understand this statement, nor the reason it has been pointed out.}). \ar{Hmmm,Yes, this is poorly expressed. I think I was carried away from some tests we did for 0.5 and 0.8, and where the photosphere was so large that the calculation of Fr would start to fail, or become sloppy, not similar to NT any more.. More varying i think I mean that for very large photospheres we risk to depart from the area of the main disk in a way.. I'm not sure if this is indeed necessary. } 

Having determined the location of the photosphere, we can now, in principle, raytrace photons from the photosphere to a far-away observer and compute the observed spectrum. Mathematically, following the notation of Ref.~\cite{Bambi:2016sac}, the observed spectrum can be written as  
\be\label{eq-thin-Fobs}
F_{\rm o} (\nu_{\rm o}) = \int I_{\rm o}(\nu_{\rm o}, X, Y) d\tilde{\Omega} \, ,
\ee
where $I_{\rm o}$ is the specific intensity detected by the far-away observer, $d\tilde{\Omega} = dX dY/D^2$ is the element of the solid angle subtended by the disk's image in the observer plane, $X$ and $Y$ are the Cartesian coordinates of the disk's image in the same plane, and $D$ is the distance between the observer and the point of emission. For computational convenience, what is done instead is to evolve the photons backwards in time from the observer to the point of emission, and to make use of a transfer function introduced in Ref.~\cite{Cunningham1975} and defined, following the notation of Ref.~\cite{Bambi:2016sac}, as
\be\label{eq-trf}
f(g^*,r_{\rm e},i) = \frac{1}{\pi r_{\rm e}} g 
\sqrt{g^* (1 - g^*)} \left| \frac{\partial \left(X,Y\right)}{\partial \left(g^*,r_{\rm e}\right)} \right| \, ,
\ee
where $r_{\rm e}$ is the emission radius, $i$ is the inclination of the observer relative to the BH spin axis,  
\be
g^* = \frac{g - g_{\rm min}}{g_{\rm max} - g_{\rm min}} \, ,
\ee
$g$ is the redshift factor, defined as the ratio of photon frequencies at observation and emission, and $g_{\rm max}$ and $g_{\rm min}$ are the maximum and minimum values of $g$ for a given $r_{\rm e}$ and $i$, respectively.

The equation for the spectrum looks like the following after the introduction of the transfer function:
%\begin{widetext}
 \be\label{eq-Fobs}
F_{\rm o} (\nu_{\rm o}) &=&
 \frac{1}{D^2} \int_{r_{\rm in}}^{r_{\rm out}} \int_0^1
\pi r_{\rm e} \frac{ g^2}{\sqrt{g^* (1 - g^*)}}\times \\ 
&& \times f(g^*,r_{\rm e},i)
I_{\rm e}(\nu_{\rm e},r_{\rm e},\vartheta_{\rm e}) \, dg^* \, dr_{\rm e} \, .
\ee
%\end{widetext}
Here, $I_{\rm e}$ is the specific intensity at the point of emission. In principle, one could calculate this by analyzing the amount of thermal radiation being upscattered by the corona redirected towards the photosphere and reprocessed there, in a self-consistent manner (as done in, e.g., Ref.~\cite{2016ApJ...826...52K}). Another possibility is to raytrace photons from the corona, modeled as a point/line on the BH spin axis within the lamppost coronal geometry model~\cite{Dauser:2013xv}, to the photosphere and reprocessed there. A third approach is to use a simple phenomenological intensity profile of a power-law (or a broken power-law). We follow the third approach and assume $I_{\rm e} \propto (r_e \sin{\Theta_{\rm top}})^{-3}$, . The calculation of the transfer function follows the scheme described in Ref.~\cite{Bambi:2016sac} with one important exception. In the present case, the reflecting surface does not always lie at the equator but at a certain height that depends on $r_e$. This dependence, moreover, is not analytical but only known numerically. We therefore implement an intersection function in the code that interpolates the disk surface using numerical data, identifies the step during evolution when this surface is crossed, and then interpolates the photon trajectory to determine the exact location where the photon would have hit the disk. 
%As Fig.~\ref{evprof} shows, in some cases the photosphere coincides with the equator at some radii. For these radii, 

Following the above scheme, we calculate iron lines for each simulation listed in Tab.~\ref{cases}, and these are plotted in Fig.~\ref{csall_IL}. We assume the disk inner edge to lie at the ISCO radius and $\dot{m}=0.1$ (results for $\dot{m}=0.3$ show the same qualitative features). For each case, we also plot the iron lines for corresponding NT disks. In all cases, the agreement between the lines from the simulated disk and the NT disk, respectively, are excellent, minor fluctuations arising from numerical inaccuracies notwithstanding. %Small differences, when they appear, appear to be numerical artifacts. 
%\co{Compare lines with different rins and routs, between simulation and NT.}
The agreement occurs at both low and high spins, and for small and large non-zero deviations. Interestingly, the deviations from the NT model seen for cases V and VI in the previous sections are not present here. Indeed, when we plot $H_{\rm ph}/r$, we find no spike in case V, unlike what we saw in plots of $H/r$. 

The above serves to validate, on the one hand, the extended \textsc{harm} code presented in this work and the additional framework to compute the reflection spectra described above, and, on the other hand, the NT-type disk based \textsc{relxill\_nk} suite for modelling the reflection spectra from accreting black holes with thin disks in the range of $\dot{m}\sim 0.1-0.3$. There is one last feature we note before closing this section. When looking at the radiated flux (Fig.~\ref{Fr}) and luminosity (Fig.~\ref{dLdr}), we find negligible difference between cases I and II, representing moderately strong gravitational fields (especially in comparison with cases IV and V representing extremely strong gravitational fields). Whereas, their iron line plots are visibly distinct, bolstering the case for x-ray reflection spectroscopy as a sensitive test of deviations from the Kerr metric.

%The convergence, once again, is very good, even for cases where the shape of the line is deformed from Kerr (cases 2, 5), which is a great success of this endeavor. These graphs are the most advanced output of this work and having such a fruitful outcome is a weighty validation of all the steps leading to this. 

\section{Future\label{sec:future}}
With advances in observational tools and techniques, higher quality x-ray data is going to be available in the coming years and decades, enabling high quality reflection spectroscopy with accreting black holes, which will provide a versatile, independent from and complimentary to other techniques, and powerful tool to perform precision tests of gravity. Precise modeling of these astrophysical systems is therefore crucial. We present a first attempt in this direction with two and three dimensional simulations of the accretion disk around a non-Kerr object described by a theory-agnostic metric. The work makes several important strides in the path towards precise tests of gravity with high fidelity:
\begin{enumerate}[label=(\alph*)]
    \item We show that accreting matter evolves in a familiar pattern and settles into a typical disk for a non-GR BH metric, as compared to typical GR BH metrics (see Fig. \ref{eveprof} in comparison to \cite{De_Villiers_2003}, \cite{2013ApJ...769..156S}, Fig. \ref{Aca}, \ref{Acb} in comparison to \cite{Schnittman_2016}).
    \item Radiated flux at source and luminosity are relatively weakly affected by deformations of the metric (being quite similar for cases I, II and III) as compared to the iron line. 
    \item We validate the approximation of the accretion disk, with $\dot{m}\sim 0.1-0.3$, to a Novikov-Thorne type razor-thin disk for modelling the reflection spectra. This was shown to be valid in the Kerr case~\cite{Penna:2010hu,Kulkarni:2011cy}, and here we show it continues to remain valid in non-Kerr cases.
\end{enumerate}
This work construes only a small step and there are many more improvements to be made in order for the framework to be of practical use: 
\begin{enumerate}[label=(\alph*)]
    \item One of the first extensions would be to perform the analysis with three-dimensional MHD. Computational constraints restricted us from performing the complete analysis in 3D, however, as we show in the Appendix, the code is ready and able to perform 3D simulations.
    \item The metric we have used is quite generic, yet it covers a subset of all possible non-Kerr metrics in the market. Other metrics, with additional interesting properties like chaos~\cite{Destounis:2021mqv}, can be implemented in the framework to see whether new features appear in the MHD evolution.
    \item More precise evaluation of the reflection spectrum requires proper modelling of the corona, by tracking thermal photons as they get reprocessed and radiated back to the disk, as done in Ref.~\cite{2016ApJ...826...52K,Kinch_2019} within GR.
    %\item \co{Anything else?}
\end{enumerate}

\section*{Acknowledgements}
S.N. acknowledges support from the Alexander von Humboldt Foundation and the Deutscher Akademischer Austauschdienst for financial support. 
The authors acknowledge support by the High Performance and Cloud Computing Group at the Zentrum f\"{u}r Datenverarbeitung of the University of T\"{u}bingen, the state of Baden-W\"{u}rttemberg through bwHPC and the German Research Foundation (DFG) through Grant No. INST 37/935-1 FUGG.

\appendix
\section{A 3D simulation}
\label{App1}

\begin{figure*}[!htb]
\begin{subfigure}{0.3\textwidth}
\includegraphics[width=1.\linewidth]{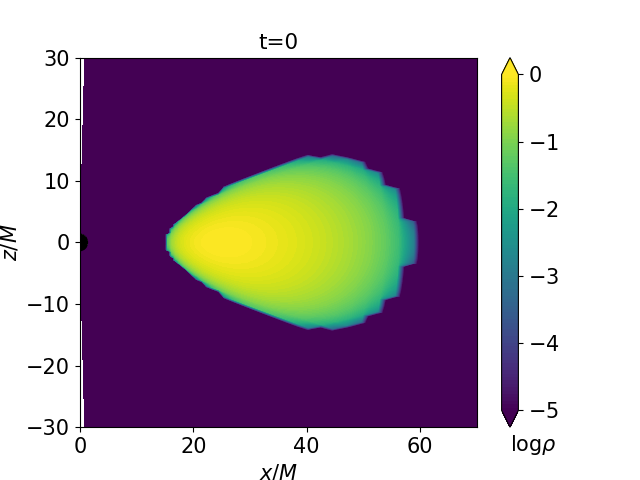}
\label{cs9_rho0}
\end{subfigure}
\begin{subfigure}{0.3\textwidth}
\includegraphics[width=1.\linewidth]{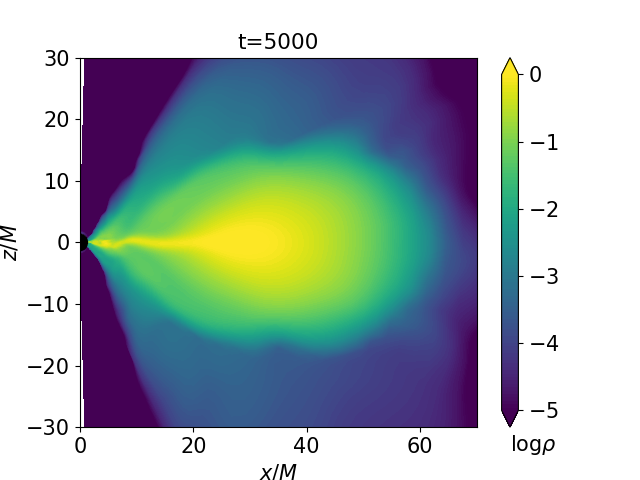}
\end{subfigure}
\begin{subfigure}{0.3\textwidth}
\includegraphics[width=1.\linewidth]{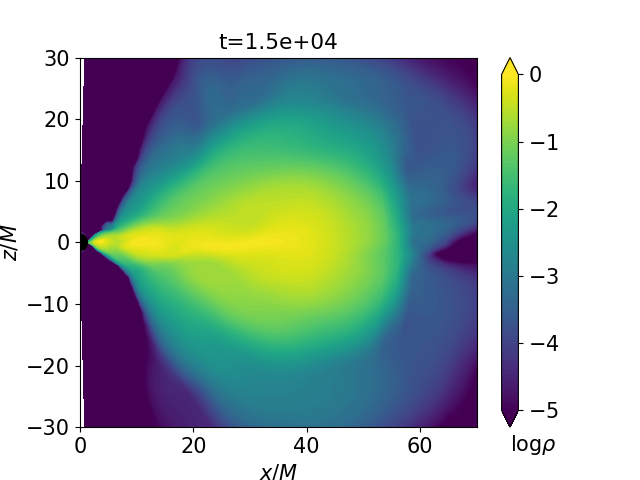} 
\end{subfigure}\\
\begin{subfigure}{0.3\textwidth}
\includegraphics[width=1.\linewidth]{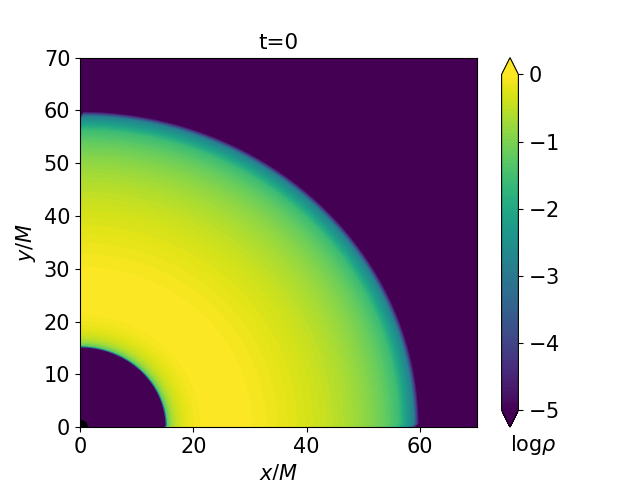}
\end{subfigure}
\begin{subfigure}{0.3\textwidth}
\includegraphics[width=1.\linewidth]{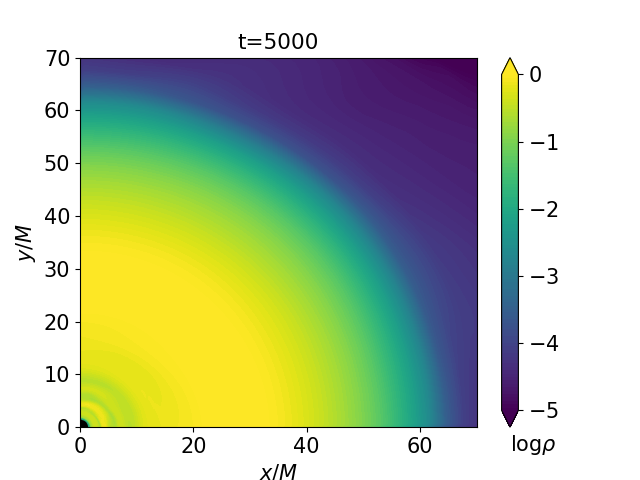} 
\end{subfigure}
\begin{subfigure}{0.3\textwidth}
\includegraphics[width=1.\linewidth]{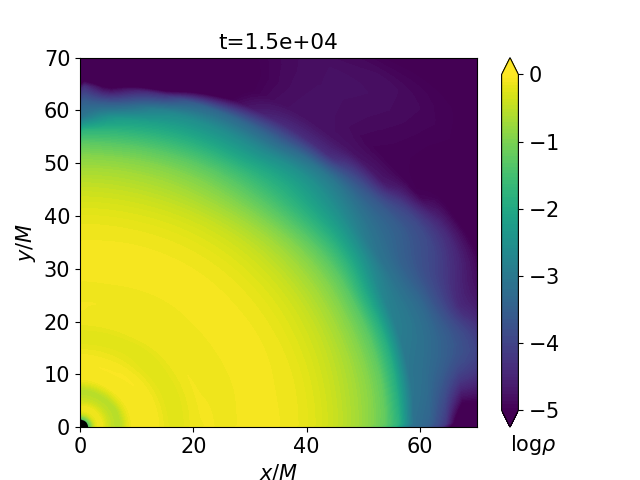}
\label{cs9_750}
\end{subfigure}
\caption{\label{fig:3d} Top: Snapshots of the disk in the $zx$-plane, i.e. slicing at a constant azimuth.\\
Bottom: Screenshots of the disk evolution in the $yx$-plane, i.e. viewing the equatorial plane from top. }
\end{figure*}

\begin{figure}[!htb]
\begin{center}
\includegraphics[width=0.33\textwidth]{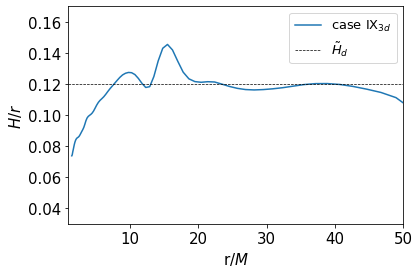} 
\caption{ Vertical thickness $H/r$, averaged over $t=8000M-15000M$, with respect to $r$. $\tilde{H}_d$ denotes the intended thickness ($=0.12$).} 
\label{cs9_hor}
\end{center}
\end{figure}

\begin{figure*}[!htb]
\begin{center}
\begin{subfigure}{0.33\textwidth}
\includegraphics[width=1\linewidth]{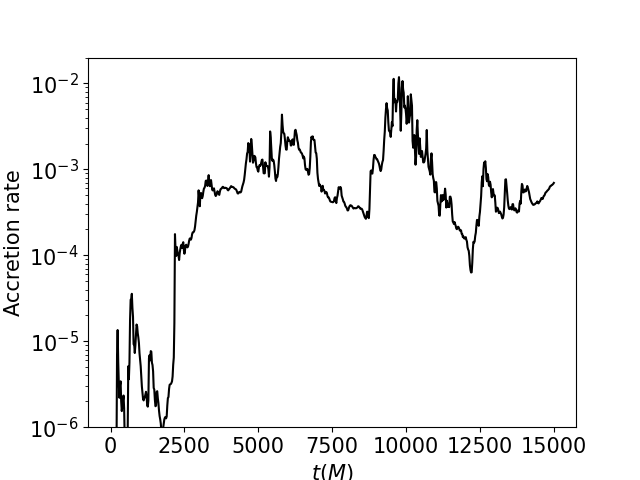} 
\end{subfigure}
\begin{subfigure}{0.33\textwidth}
\includegraphics[width=1\linewidth]{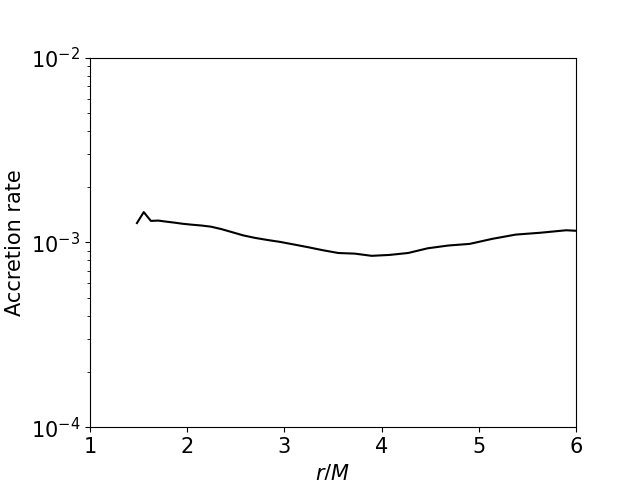} 
\end{subfigure}
\caption{ Top: Accretion rate at the horizon, as a function of time.
Bottom: Time averaged accretion rate, in units of $M_{\rm code}/M$ and averaged over $t= 8000M - 15000M$, with respect to $r$.}
\label{cs9_Ac}
\end{center}
\end{figure*}

\begin{figure}[!htb]
\begin{center}
\includegraphics[width=0.33\textwidth]{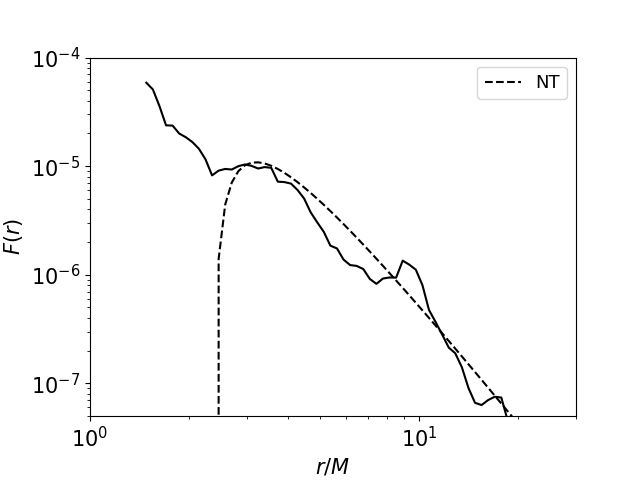} 
\caption{ Radiated flux per unit area in the fluid frame, averaged over $t= 8000M - 15000M$, as a function of radius $r$. The dashed line denotes the corresponding line as predicted from the NT model, and described in the text.} 
\label{cs9_Fr}
\end{center}
\end{figure}

Ideally, we would like to have carried out all simulations in a 3D configuration. Although that was not possible due to computational costs and time limitations, we did perform one simulation in 3D to make sure that the code is capable of running in 3D and to compare the results with the 2D simulations. Note that the resolution is halved in the ($r$, $\theta$) directions than the 2D simulations, while the resolution in the azimuthal direction is set to 64  (i.e. $96\times 96\times 64$). The simulation parameters are listed in Tab.~\ref{cases} under case $9\_3$D, and the results are shown in a series of plots in Figs.~\ref{fig:3d}-\ref{cs9_Fr}. 

Fig.~\ref{fig:3d} shows the disk at various epochs (initial, intermediate, final). The initial disk is the same as in the 2D simulations (see Sec.~\ref{sec:initialdisk} for a description of the initial disk profile); in particular, it starts off  independent of the azimuthal coordinate. Over time, the disk acquires a 3D profile, as can be seen in the bottom set of plots. Still, the 2D cross-section at a constant azimuth shows, in the top set of plots, a profile similar to the 2D simulations (cf. Fig.~\ref{eveprof}). 

Fig.~\ref{cs9_hor} shows the vertical thickness $H/r$, which, as in the 2D simulations (Fig.~\ref{Hor}), shows that desired thickness ($0.12$) was achieved. Fig.~\ref{cs9_Ac} presents the accretion rate, as a function of time (left panel) at the horizon and as a function of radius (right panel) averaged over $t=8000M-15000M$. In both cases, the accretion rate is along expected lines (cf. Figs.~\ref{Aca} and~\ref{Acb}). Finally, in Fig.~\ref{cs9_Fr} we show the radiated flux from the disk (cf. Fig.~\ref{Fr}). Outside the ISCO, there is very good agreement between the flux calculated with the 3D simulation and the flux from the analytical model, as was the case with 2D simulations. 

%The way the results are shown does not differ from the main results section. In figures \ref{cs9_rho0}-\ref{cs9_750} screenshots from the disk evolution in different planes are visible. Next, in figure\ref{cs9_hor} the averaged density scale height $(H/r)$ is shown. Overall the disk structure is decent, with a small bump in the thickness appearing, probably due to low resolution. 

%In figure \ref{cs9_Ac} we have the accretion rate with respect to time and $r$. Despite the low resolution, the accretion with respect to $r$ appears very stable. It has been mentioned already that accretion rate is better resolved with the inclusion of the third dimension. 

%Lastly, in figure \ref{cs9_Fr}, the locally radiated flux with respect to $r$ is shown. The line looks to be in good compliance with the NT prediction.   

%\bibliographystyle{plain}
\bibliography{references}
\end{document}